\documentclass[apjl]{emulateapj}
\usepackage{amsfonts,amsmath,graphicx,natbib,apjfonts,subfigure,textcomp,color}

\def\nyu{1}
\def\cfa{2}
\def\neth{3}
\def\ou{4}
\def\arcetri{5}
\def\toronto{6}
\def\southa{7}
\def\kavli{8}
\def\michigan{9}
\def\az{10}
\def\pasadena{11}
\def\ferrara{12}
\def\goddard{13}
\def\pasadenabis{14}

\shorttitle{The Supernova Chameleon SN\,2014C}
\shortauthors{Margutti et al.}

\begin{document}
\title{Ejection of the massive hydrogen-rich envelope timed with the collapse of the stripped SN\,2014C}

\author{Raffaella Margutti\altaffilmark{\nyu,\cfa}, A. Kamble\altaffilmark{\cfa}, D. Milisavljevic\altaffilmark{\cfa}, S. de Mink\altaffilmark{\neth}, E. Zapartas\altaffilmark{\neth}, M. Drout\altaffilmark{\cfa}, R. Chornock\altaffilmark{\ou}, G. Risaliti\altaffilmark{\arcetri}, B.~A. Zauderer\altaffilmark{\cfa}, M. Bietenholz\altaffilmark{\toronto,\southa}, M. Cantiello\altaffilmark{\kavli}, S. Chakraborti\altaffilmark{\cfa}, L. Chomiuk\altaffilmark{\michigan}, W. Fong\altaffilmark{\az}, B. Grefenstette\altaffilmark{\pasadena}, C. Guidorzi\altaffilmark{\ferrara}, R. Kirshner\altaffilmark{\cfa},  J.~T. Parrent\altaffilmark{\cfa}, D. Patnaude\altaffilmark{\cfa}, A.~M. Soderberg\altaffilmark{\cfa},  N.~C. Gehrels\altaffilmark{\goddard}, F. Harrison\altaffilmark{\pasadenabis}}

\altaffiltext{\nyu}{Center for Cosmology and Particle Physics, New York University, 4 Washington Place, New York, NY 10003, USA}
\altaffiltext{\cfa}{Harvard-Smithsonian Center for Astrophysics, 60 Garden St., Cambridge, MA 02138, USA}
\altaffiltext{\neth}{Anton Pannenkoek Institute for Astronomy, University of Amsterdam, 1090 GE Amsterdam, The Netherlands}
\altaffiltext{\ou}{Astrophysical Institute, Department of Physics and Astronomy, 251B Clippinger Lab, Ohio University, Athens, OH 45701, USA}
\altaffiltext{\arcetri}{INAF-Arcetri Astrophysical Observatory, Largo E. Fermi 5, I-50125 Firenze, Italy}
\altaffiltext{\toronto}{Department of Physics and Astronomy, York University, Toronto, ON M3J 1P3, Canada}
\altaffiltext{\southa}{Hartebeesthoek Radio Observatory, PO Box 443, Krugersdorp 1740, South Africa}
\altaffiltext{\kavli}{Kavli Institute for Theoretical Physics, University of California, Santa Barbara, CA 93106, USA}
\altaffiltext{\michigan}{Department of Physics and Astronomy, Michigan State University, East Lansing, MI 48824, USA}
\altaffiltext{\az}{Steward Observatory, University of Arizona, 933 North Cherry Avenue, Tucson, AZ 85721, USA}
\altaffiltext{\pasadena}{Cahill Center for Astrophysics, 1216 E. California Blvd., California Institute of Technology, Pasadena, CA 91125, USA}
\altaffiltext{\ferrara}{University of Ferrara, Department of Physics and Earth Sciences, via Saragat 1, IÐ44122 Ferrara, Italy}
\altaffiltext{\goddard}{NASA Goddard Space Flight Center, Code 661, Greenbelt, MD 20771, USA}
\altaffiltext{\pasadenabis}{Space Radiation Laboratory, California Institute of Technology, 1200 E California Blvd, MC 249-17, Pasadena, CA 91125, USA}

\begin{abstract}
We present multi-wavelength observations of Supernova 2014C during the first 500 days of its evolution. These observations represent the first solid detection of a young extragalactic stripped-envelope SN out to high-energy X-rays $\sim$ 40 keV. SN\,2014C was the explosion of an envelope-stripped progenitor star with ordinary explosion parameters ($E_{\rm{k}}\sim1.8\times10^{51}\,\rm{erg}$ and $M_{\rm{ej}}\sim1.7\,\rm{M_{\sun}}$). However, over the time scale of $\sim$1 yr, SN\,2014C experienced a complete metamorphosis and evolved from an ordinary hydrogen-poor supernova  of type Ib into a strongly interacting, hydrogen-rich supernova of type IIn, thus violating the traditional classification scheme of type-I vs. type-II SNe. Signatures of the SN shock interacting with a dense medium are observed across the electromagnetic spectrum, from the radio to the hard X-ray band. Coordinated observations with Swift, Chandra and NuSTAR have captured the evolution in detail and revealed the presence of a massive shell of $\sim1\,\rm{M_{\sun}}$ of hydrogen-rich material at $\sim6\times10^{16}\,\rm{cm}$ from the explosion site. We estimate that the shell was ejected by the progenitor star in the decades to centuries before core collapse. This result poses significant challenges to current theories of massive star evolution, as it requires a physical mechanism responsible for the ejection of the deepest hydrogen layer of H-poor SN progenitors synchronized with the onset of stellar collapse. Theoretical investigations point at  binary interactions and/or instabilities during the last stages of nuclear burning in massive stars as potential triggers of the highly time-dependent mass loss. We constrain these scenarios utilizing the sample of 183 SNe Ib/c with public radio observations. Our analysis identifies SN\,2014C-like signatures, consistent with strong interaction, in $\sim$$10$\% of SNe with constraining radio data. This fraction is somewhat larger but reasonably consistent with the expectation from the theory of recent envelope ejection due to binary evolution \emph{if} the ejected material can survive in the close environment for $10^3-10^4$ yrs. Alternatively, nuclear burning instabilities extending all the way to the core C-burning phase might also play a critical role.
\end{abstract}
\keywords{supernovae: specific (SN\,2014C)}
\section{Introduction}
\label{Sec:Intro}

Mass loss from massive stars ($>10\,\rm{M_{\sun}}$) plays a major role in the chemical enrichment of the Universe and directly determines the luminosity,  lifetime and fate of stars. Yet, the dominant channels and the physical mechanisms that drive mass loss in evolved massive stars are uncertain (see \citealt{Smith14} for a recent review). This lack of understanding is significant as it further impacts our estimates of the stellar initial mass function in galaxies and star formation through cosmic time, which rely on the predictions of stellar evolution models (\citealt{Bastian10}, \citealt{Madau98}, \citealt{Hopkins06}).


Relevant observations that expose our inadequate theoretical understanding of mass loss in evolved massive stars include the discovery of  powerful eruptions prior to major explosions in H-rich massive stars, of which SN\,2009ip is the best studied example (e.g. \citealt{Mauerhan13}, \citealt{Pastorello13}, \citealt{Margutti14}, \citealt{Smith14} and references therein; \citealt{Ofek13}, \citealt{Ofek14}; \citealt{Fraser13}). 
A precursor to the SN explosion has also been identified in the case of the type-Ibn SN\,2006jc, which showed signs of interaction with a He-rich  medium (e.g. \citealt{Foley07}, \citealt{Pastorello07}).
Evidence for significantly enhanced mass loss timed with the explosion has been found for the H-poor progenitors of both type-IIb SNe (\citealt{Kamble15}, \citealt{Maeda15}) and type-Ib SNe (\citealt{Svirski14}), as well as for the H- and He-poor progenitors of type-Ic SNe associated with some nearby gamma-ray bursts (\citealt{Margutti15}, \citealt{Nakar15}).
Along the same line, it is relevant to mention the possible detection of an outburst from the progenitor of the  type-Ic SN PTF11qcj $\sim2.5$ yrs before stellar death (\citealt{Corsi14}), the evidence for significant temporal variability in the radio light-curves of Ib/c SNe (\citealt{Soderberg07}, \citealt{Wellons12}) and  the recent detection of interaction of the  H-poor super-luminous SN iPTF13ehe with H-rich material at late-times (\citealt{Yan15}). 

These observations point to the presence of strongly time-dependent mass loss synchronized with core-collapse in a variety of stellar explosions (from type-IIn SNe to ordinary Ib/c, gamma-ray burst SNe and even super-luminous SNe). 
The erratic behavior of these stars approaching stellar death across the mass spectrum clearly deviates from the commonly accepted picture of steady mass loss through line-driven winds employed by current models of stellar evolution (e.g. \citealt{Smith14}). However, the nature of the physical process responsible for the highly time-dependent mass loss  is  at the moment a matter of debate. 
Equally unclear is the extent to which these processes have an active and important role in the evolutionary path that leads to the envelope-stripped progenitors of ordinary hydrogen-poor core-collapse SNe (i.e. SNe of type Ib/c). We address this still-open question in our study of SN\,2014C in the context of 183 Ib/c radio SNe.


We present multi-wavelength observations of the remarkable metamorphosis of SN\,2014C, which evolved from an ordinary H-stripped core-collapse SN of type Ib into a strongly interacting type-IIn SN over $\sim1$ yr of observations. The relative proximity of SN\,2014C in  NGC\,7331 (d=14.7 Mpc, \citealt{Freedman01}) allowed us to witness the progressive emergence of observational signatures of the undergoing interaction across the electromagnetic spectrum, and, in particular, it offered us the unprecedented opportunity to follow the development of luminous X-ray emission captured in detail by the Swift X-ray Telescope (XRT), the Chandra X-ray Observatory (CXO) and the Nuclear Spectroscopic Telescope Array (NuSTAR). SN\,2014C is the first young H-stripped SN for which we have been able to follow the evolution in the hard X-ray range. SN\,2014C is also the first envelope-stripped SN that showed a mid-InfraRed re-brightening in the months after the explosion \citep{Tinyanont16}.

This paper is organized as follows. UV, optical, X-ray and hard X-ray observations of SN\,2014C are described in Sec. \ref{Sec:obs}. The explosion properties of SN\,2014C are derived in Sec. \ref{SubSec:ExplPar}. We interpret the metamorphosis of SN\,2014C in the context of strong SN shock interaction with dense and massive H-rich material that was ejected by the progenitor star during the years before core-collapse. We derive mass, density, distance and time of ejection of the H-rich shell  in Sec. \ref{Sec:Env}. We put SN\,2014C into the context of Ib/c SNe and constrain the rate of 14C-like events among 183 Ib/c SNe observed at radio frequencies in Sec. \ref{Sec:Rates}. Based on these results, we discuss the challenges faced by the current theories of massive star evolution and explore alternatives in Sec. \ref{Sec:int}. We conclude in Sec. \ref{Sec:conc}. Details of the spectroscopic evolution of SN\,2014C are provided in \cite{Milisavljevic15} (hereafter M15), while we refer to Kamble et al., 2015 (hereafter K15) for the modeling of the radio synchrotron emission which originates from the SN shock interaction with the medium.

Accurate modeling of the bolometric light-curve of SN\,2014C of Sec. \ref{SubSec:ExplPar} constrains the time of first light to December 30, 2013 $\pm1$ day (MJD 56656 $\pm1$, see Sec. \ref{SubSec:ExplPar}). Times will be referred to MJD 56656 unless explicitly noted. M15 estimate  $E(B-V)_{tot}\sim0.75$ mag in the direction of the transient, which we will use to correct our photometry.  The Galaxy only contributes a limited fraction, corresponding to $E(B-V)_{\rm{mw}}=0.08$ mag  \citep{Schlafly11}.  Local extinction thus clearly dominates over the Galactic value, suggesting the presence of large quantities of material in NGC\,7331 towards SN\,2014C.  Finally, uncertainties are quoted at the level of $1\,\sigma$  confidence level unless stated otherwise. 

\section{Observations and data reduction}
\label{Sec:obs}

\subsection{Optical-UV Photometry with Swift-UVOT}

\begin{figure}
\vskip -0.0 true cm
\centering
\includegraphics[scale=0.35]{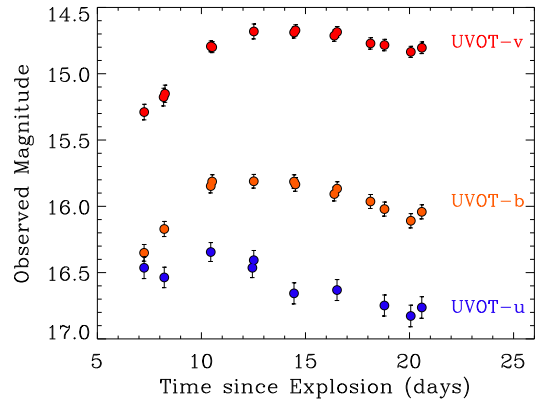}
\caption{Temporal evolution of SN2014C in $u$, $b$ and $v$ band as captured by Swift-UVOT. No host subtraction and extinction correction have been applied.}
\label{fig:UVOT}
\end{figure}

\begin{figure}
\vskip -0.0 true cm
\centering
\includegraphics[scale=0.56]{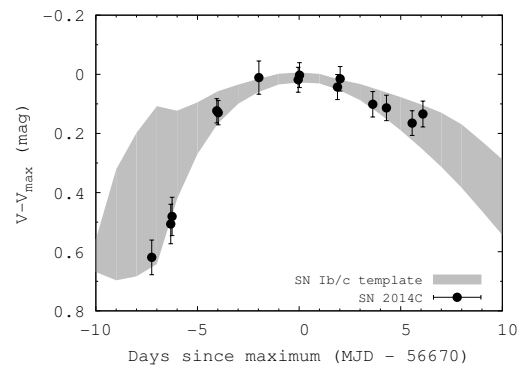}
\caption{Temporal evolution of the $v$-band emission from SN\,2014C as observed by \emph{Swift}-UVOT, compared to the SNe Ibc template from   \cite{Drout11}. The width of the grey curve is derived from the 1$\sigma$ deviation from the mean at each epoch.}
\label{fig:lc}
\end{figure}


The UV-Optical Telescope (UVOT, \citealt{Roming05}) onboard the \emph{Swift} satellite \citep{Gehrels04} started observing SN\,2014C on January 6, 2014 (PIs P. Milne and R. Margutti). Due to its angular proximity to the Sun, SN\,2014C was not observable by \emph{Swift} in the time periods late January-April 2014 and February-March 2015. We employed the latest HEAsoft release (v. 6.16) with the corresponding calibration files to reduce the data and a source extraction region of $3"$ to minimize the contamination from host-galaxy light. We extracted the photometry following the prescriptions by \cite{Brown09}. 

SN\,2014C is clearly detected in the wave-length range 3500-5500 \AA\, (i.e. UVOT $u$, $b$ and $v$ filters) between $-7$ days and $+7$ days since maximum light (MJD 56663 $<$t$<$ MJD 56677). In the same time period SN\,2014C is only marginally detected in the UV (i.e. UVOT $w1$, $w2$ and $m2$ filters), mainly because of bright and diffuse UV emission from the host galaxy at the location of the SN. SN\,2014C reaches v-band maximum light around January 13, 2014 (MJD 56670). A search for increased UV emission arising from the SN shock interaction with the massive CSM in late-time ($t>6$ months) UVOT data led to a non-detection. Table \ref{Tab:UVOTphot} reports the complete UVOT photometry. 

A comparison of the $v$-band light-curve of SN\,2014C to the SN Ibc template by \cite{Drout11} in Figure \ref{fig:lc} illustrates that its rise time to maximum light of $\sim14$ days falls on the short end of the observed distribution. The overall shape of the light-curve is however in reasonable agreement with the template, which indicates a rather standard ejecta mass ($M_{\rm{ej}}$) to kinetic energy ($E_{\rm{k}}$) ratio, as quantified in Sec. \ref{SubSec:ExplPar}. 


\subsection{Deep Late-time Optical Photometry with MMTCam}
We obtained $r'i'-$band observations of SN\,2014C with the MMTCam imager mounted on the 6.5m MMT on 5 epochs spanning May 18, 2014  to May 23, 2015 (139 to 509 days since explosion, PI Margutti).  All frames were bias, dark, and flat field corrected using standard tasks in IRAF\footnote{IRAF is distributed by the National Optical Astronomy Observatory, which is operated by the Association for Research in Astronomy, Inc.\, under cooperative agreement with the National Science Foundation.}.  PSF photometry was performed on all images and absolute calibration was performed using five SDSS stars in a nearby standard field.  The resulting photometry is listed in Table \ref{Tab:MMT} and shown in Figure \ref{Fig:XRO}.  No template subtraction was performed as the source was still visible in our final epoch of observation, and no sufficiently deep archival images were available in these bands.  During our final epochs of observation the source is $\sim$ 0.6 mag brighter than the R-band pre-explosion source described in M15.
\subsection{Early-time X-ray observations with Swift-XRT}
\label{SubSec:XRT}
The \emph{Swift} X-Ray Telescope (XRT, \citealt{Burrows05}) started observing SN\,2014C on  ($\delta t\sim7$ days since explosion, PI P. Milne).  Observations acquired before SN\,2014C set behind the Sun cover the time interval $ t\sim7-20$ days since the explosion, with a total exposure time of $17.2$ ks. We detect significant X-ray emission at the SN site. However, inspection of pre-explosion images acquired in 2007  reveals the presence of diffuse X-ray emission that is not spatially resolved by the XRT. By accounting for the unresolved host-galaxy contribution, as constrained from pre-explosion observations, we infer a 3$\sigma$ limit to the SN emission of  $8.1\times 10^{-4}\,\rm{c\,s^{-1}}$ (0.3-10 keV).   

Coordinated observations of the Chandra X-ray Observatory and NuSTAR obtained at later times (Sec. \ref{SubSec:ChandraData} and \ref{SubSec:NuSTAR}) showed evidence for large intrinsic neutral hydrogen absorption in the direction of SN\,2014C. At the time of the first Chandra observations we infer a total hydrogen column density $\rm{NH_{tot}}\lesssim4\times10^{22}\,\rm{cm^{-2}}$ (Sec. \ref{SubSec:CXONuSTAR}). The Galactic hydrogen column density is $\rm{NH_{mw}}\lesssim6.1\times10^{20}\,\rm{cm^{-2}}$ \citep{Kalberla05}. The measured hydrogen column is thus dominated by material in the host galaxy of SN2014C. Restricting our analysis  to the 2-10 keV energy range to minimize the impact of the uncertain absorption of soft X-rays, and accounting for the unresolved host-galaxy contribution, we infer a 3$\sigma$ limit to the SN emission of  $3.0\times 10^{-4}\,\rm{c\,s^{-1}}$ (2-10 keV).
For $\rm{NH_{tot}}\sim 5\times10^{22}\,\rm{cm^{-2}}$, the corresponding unaborbed flux limit is
$F_x<4.1\times 10^{-14}\,\rm{erg\,s^{-1}cm^{-2}}$ and the luminosity limit is $L_x<1.1\times 10^{39}\,\rm{erg\,s^{-1}}$
(2-10 keV). We assume a power-law spectral model with photon index $\Gamma=2$, as appropriate for non-thermal Inverse Compton (IC) emission (see Sec. \ref{Sec:Env}).

\subsection{Deep X-ray observations with Chandra}
\label{SubSec:ChandraData}

\begin{figure}
\vskip -0.0 true cm
\centering
\includegraphics[scale=0.35]{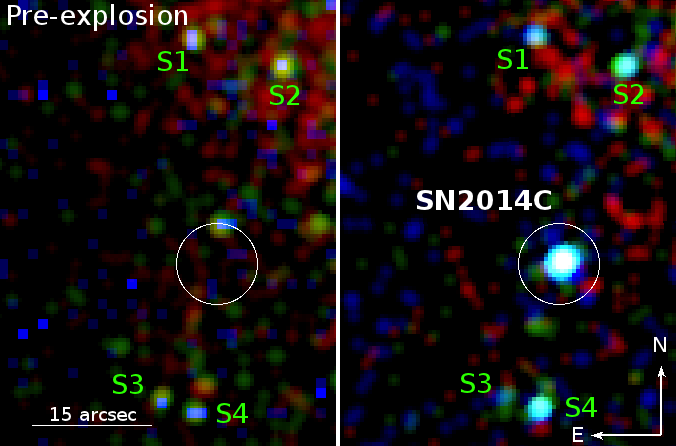}
\caption{Pre- and post-explosion, false-color composite X-ray images at the location of SN\,2014C taken with Chandra. In both panels we use red  are for for the 0.3-1 keV energy band, green for 1-3 keV photons while blue-to-white shades mark the hardest photons in the images with energy 3-10 keV. The pre-explosion image collects 29.5 ks of observations acquired in 2001. The right panel collects the post-explosion Chandra data presented in this paper (exposure time of 29.7 ks), covering the time period November 2011- April 2015.  SN\,2014C is well detected in this time period as a bright source of hard X-ray emission. White circle: $5"$ radius region at the position of SN\,2014C.}
\label{fig:prepostXrays}
\end{figure}

\begin{figure}
\vskip -0.0 true cm
\centering
\includegraphics[scale=0.6]{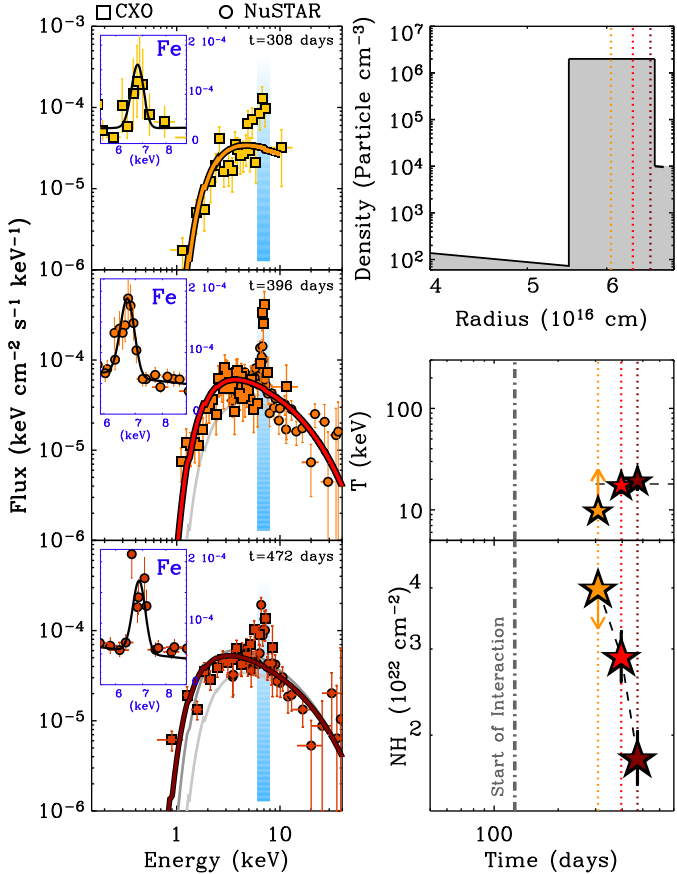}
\caption{Coordinated CXO (squares) and NuSTAR (filled dots) observations of SN\,2014C revealed an X-ray  thermally emitting plasma with characteristic temperature $T\sim20$ keV and an absorption decreasing with time (middle and bottom panels on the right). The best fitting bremsstrahlung model is represented with a thick colored line in each panel on the left and reproduced with grey lines in the other panels for comparison. An excess of emission around 6.7-6.9 keV is clearly detected at all epochs. We associate this emission with He-like and H-like Fe transitions. The upper right panel potrays the density profile of the environment as constrained by these observations and our modeling in Sec. \ref{Sec:Env}. }
\label{Fig:ChandraNustar}
\end{figure}

\begin{figure}
\vskip -0.0 true cm
\centering
\includegraphics[scale=0.4]{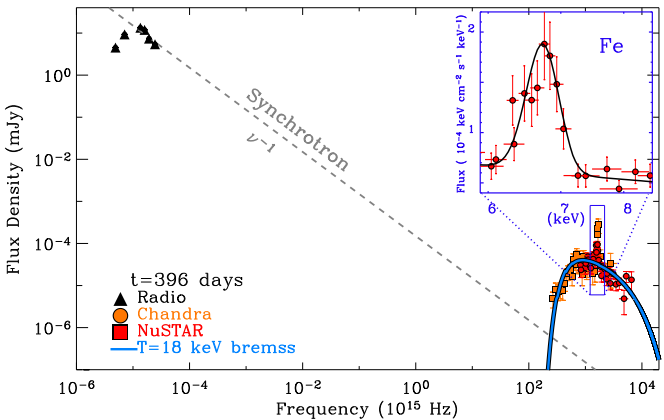}
\caption{Radio (VLA from K15) to hard X-ray (CXO, NuSTAR) spectral energy distribution of SN\,2014C at $t=396$ days after the explosion, shown here as an example. The X-ray emission is in clear excess to the synchrotron model that best fits the radio observations, as expected in the case of SN shock interaction with a very dense medium (e.g. \citealt{Chevalier06}). CXO and NuSTAR data are best fit by an absorbed bremsstrahlung model with $T\sim 18$ keV and $\rm{NH_{tot}\sim 3\times 10^{22}\,\rm{cm^{-2}}}$. Emission at 6.7-6.9 keV due to H-like and He-like Fe  transitions is also clearly detected (\emph{Inset}).}
\label{Fig:SED}
\end{figure}

\emph{Pre-Explosion:} The field of SN\,2014C was observed by the Chandra X-ray Observatory (CXO) on January 27th, 2001 (PI Zezas, ID 2198). In 29.5 ks of observations we find no evidence for X-ray emission at the SN site down to the limit of $2.6\times 10^{-4}\,\rm{c\,s^{-1}}$ (0.3-10 keV). For an assumed power-law spectrum with photon index $\Gamma=2$ the value above converts into an absorbed flux $<2.8\times10^{-15}\,\rm{erg\,s^{-1}\,cm^{-2}}$ (0.3-10 keV), which corresponds to $L_x<8.5\times10^{37}\,\rm{erg\,s^{-1}}$  (0.3-10 keV). 

\emph{Post-Explosion:} We started a monitoring campaign with the CXO to constrain the evolution of SN\,2014C under our approved program to monitor SNe originating from hydrogen stripped progenitors in the local Universe. A first observation was obtained on November 3, 2014 ($t=308$ days, PI A. Soderberg, ID 16005), followed by a number of observations during 2015 under a multi-epoch DDT program (PI R. Margutti, IDs 17569, 17570, 17571).

\begin{figure}
\vskip -0.0 true cm
\centering
\includegraphics[scale=0.4]{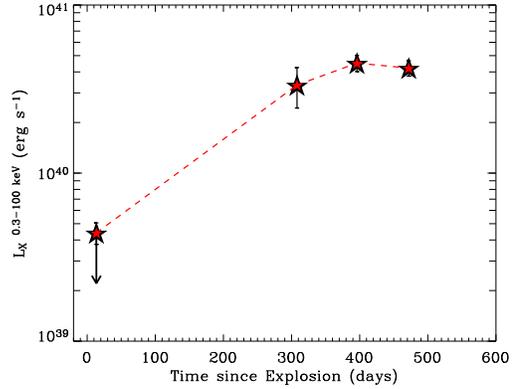}
\caption{Broad-band X-ray light-curve of SN\,2014C during the first 500 days as captured by Swift, the CXO and NuSTAR.}
\label{Fig:XrayBol}
\end{figure}

CXO data have been reduced with the CIAO software package (version 4.6) and corresponding calibration files. Standard ACIS data filtering has been applied. SN\,2014C is a bright source of X-ray emission with luminosity increasing  with time (Fig. \ref{fig:prepostXrays} and Fig. \ref{Fig:XrayBol}).  In our first epoch at $t=308$ days, SN\,2014C is detected at the level of $>40\,\sigma$, with a net count-rate of  $0.013\,\rm{c\,s^{-1}}$ (0.5-8 keV, exposure time of 9.9 ks). A spectral fit with an absorbed power-law model $F_{\nu}\propto \nu^{-\beta}$ indicates very hard emission with index $\beta=-1.5\pm0.3$.  The detected X-ray emission is also in significant excess with respect to the extrapolation of the synchrotron spectrum that best fits the radio observations at all times (Fig. \ref{Fig:SED}). These two findings together suggest a thermal origin for the X-rays (as confirmed by coordinated CXO-NuSTAR observations obtained later).  A fit with an absorbed  bremsstrahlung model constrains $T>10$ keV. We estimate the absorption at $t=308$ days as follows. Chandra and NuSTAR observations obtained $\sim90$ days later, at $t=396$ days, are well modeled by a thermal bremsstrahlung spectrum with  $T\sim18$ keV (Sec. \ref{SubSec:CXONuSTAR}). The interaction of the SN shock with a very dense medium causes a rapid deceleration of the forward shock accompanied by a sudden and marked increase of the reverse shock temperature. At later times the temperatures of the forward and reverse shock decrease (see e.g. \citealt{Chugai06}, their Fig 2). SN\,2014C started to interact with the dense shell $\sim100$ days after the explosion (M15, i.e. $\sim200$ days before the coordinated CXO-NuSTAR follow up, Sec. \ref{Sec:Env}), which implies $T>18$ keV at $t=308$ days. Using this constraint to the temperature in our absorbed bremsstrahlung fit to the CXO data, we infer $\rm{NH_{tot}}\lesssim 4\times10^{22}\,\rm{cm^{-2}}$ at $t=308$ days.

Ninety days later, ($t=396$ days), our campaign reveals that SN\,2014C substantially brightened with time, reaching $0.026\,\rm{c\,s^{-1}}$ in the 0.5-8 keV band ($>90\,\sigma$ significance level detection using 9.9 ks of observations). From the spectral analysis it is clear that the re-brightening is more apparent at soft X-ray energies ($E\lesssim4$ keV), pointing to a decreased neutral hydrogen column density $\rm{NH}_{\rm{tot}}$. 
Our latest CXO observation was obtained at $t=472$ days since explosion, with total exposure of 9.9 ks. SN\,2014C is detected at the level of $0.0285\,\rm{c\,s^{-1}}$ ($>100\,\sigma$ significance level, 0.5-8 keV).  The spectral parameters at $t=396$ days and $t=472$ days are best constrained by the joint CXO-NuSTAR fit described in Sec. \ref{SubSec:CXONuSTAR}. The results from our broad band X-ray spectral fits and the resulting luminosities are reported in Table \ref{Tab:Spec}. Finally, in each of the three CXO observations we note the presence of enhanced emission around 6.7-6.9 keV that we associate with H-like and He-like Fe line emission (Fig. \ref{Fig:ChandraNustar}).


\subsection{Hard X-ray observations with NuSTAR}
\label{SubSec:NuSTAR}

We obtained two epochs of observations with the Nuclear Spectroscopic Telescope Array (NuSTAR) under approved DDT and Guest Investigator programs (PI Margutti), coordinated in time with the CXO at $t=396$ days and $t=472$ days since explosion. Our programs led to the first detection of a young hydrogen-stripped core-collapse SN at hard X-rays energies. SN\,2014C is well detected by NuSTAR in the energy range $3-40$ keV. The NuSTAR level 1 data products have been processed with the NuSTAR Data Analysis Software package version 1.4.1 included in the 6.16 HEASoft release. Event files were produced, calibrated, and cleaned using the standard filtering criteria and the latest files available in the NuSTAR calibration database (CALDB version 20150622). The total net exposures, after the exclusion of periods of source occultation by Earth and passages on the South Atlantic Anomaly, are 32.5 and 22.4 ks  for the first and second observation, respectively. The source extraction radius is 1$'$ for both observations, and has been chosen in order to maximize the S/N ratio. The background has been extracted in source-free regions in the the field of view of each observation. The NuSTAR data are calibrated up to 79 keV, however a comparison between the source and the background counts show that the spectrum is background-dominated above 30-40 keV. Therefore we limited our spectral analysis to the range 3-40 keV.
The higher energy bandpass of NuSTAR and the joint observations with the CXO at lower energies are crucial to constrain the temperature of the emitting region, the absorption column density of material in front of the emitting region and their evolution with time (Sec. \ref{SubSec:CXONuSTAR}). 

\subsection{CXO-NuSTAR spectral modeling}
\label{SubSec:CXONuSTAR}

\begin{deluxetable}{cccc}[h!]
\tabletypesize{\scriptsize}
\tablecolumns{4} 
\tablewidth{20pc}
\tablecaption{Broad-band X-ray spectral modeling with thermal bremsstrahlung}
\tablehead{  \colhead{Date} & \colhead{Instrument} & \colhead{Temperature}& \colhead{Absorption}\\
(days) & & $T$ (keV) & $\rm{NH_{tot}}$ ($10^{22}\,\rm{cm^{-2}}$)}
\startdata
308 & CXO & $>10$ & $\lesssim 4$\\
396 & CXO+NuSTAR & $17.8 ^{+3.7}_{-2.8}$ & $2.9^{+0.4}_{-0.3}$ \\
472 &CXO+NuSTAR & $19.8^{+6.3}_{-3.9} $ & $1.8^{+0.2}_{-0.2}$ \\
\enddata
\label{Tab:Spec}
\end{deluxetable}

\begin{deluxetable}{ccccc}[h!]
\tabletypesize{\scriptsize}
\tablecolumns{4} 
\tablewidth{20pc}
\tablecaption{Properties of the Fe line emission modeled with a Gaussian profile}
\tablehead{  \colhead{Date} & \colhead{Instrument} & \colhead{Central Energy}& \colhead{FWHM}& \colhead{Flux}\\
(days) & & $E$ (keV) & $(\rm{keV})$ &($10^{-13}\,\rm{erg\,s^{-1}\,cm^{-2}}$)}
\startdata
308 & CXO & $6.80\pm0.20$ & $0.55\pm0.23$ & $(1.30\pm0.30)$\\
396 & CXO+NuSTAR & $6.70\pm0.04$ &  $0.56\pm 0.09$ &  $(1.20\pm0.10)$\\
472 &CXO+NuSTAR & $6.84\pm0.05$ &  $0.59\pm 0.14$ & $(1.20\pm0.10)$ \\
\enddata
\label{Tab:Fe}
\end{deluxetable}

The CXO covers the energy range $0.3-10$ keV, while NuSTAR is sensitive between $3$ and $79$ keV. The two instruments have very different Point Spread Functions (PSF): while the CXO is able to spatially resolve the emission from SN\,2014C from other sources in NGC7331 (Fig. \ref{fig:prepostXrays}), the composite emission appears as a single source at higher energies due to the wider instrumental PSF of NuSTAR (FWHM of 18"). The emission from other sources within the NuSTAR $1'$  region is significantly fainter than SN\,2014C. Nevertheless, to estimate and remove the contamination from other sources to the NuSTAR PSF we employed the CXO observations as follows. For both epochs we extracted a CXO spectrum of the contaminating sources by using an annular region with inner radius $1.5''$ and outer radius of $1'$ centered at the SN position. We model this spectrum with an absorbed power-law model to determine the best fitting spectral parameters of the contaminating emission, and extrapolate its contribution to the NuSTAR energy band. We then add a spectral component with these parameters to the model used for the spectral fitting of the NuSTAR data, only. As a refinement of the method above, we extracted a spectrum of each point-like source that we detected with the CXO within the NuSTAR extraction region and fit the spectrum of each source with an absorbed power-law function that we extrapolate to the NuSTAR energy band, obtaining consistent results. 

Accounting for the contaminating emission to the NuSTAR data as described above, we find that the two epochs of coordinated CXO-NuSTAR observations are well fit by an absorbed thermal bremsstrahlung spectral model with temperature $T\sim20$ keV and decreasing absorption with time (Fig. \ref{Fig:ChandraNustar}). We measure  $\rm{NH_{tot}}\sim 3\times10^{22}\,\rm{cm^{-2}}$ and $\rm{NH_{tot}}\sim 2\times10^{22}\,\rm{cm^{-2}}$ at $t=396$ days and $t=472$ days, respectively. Table  \ref{Tab:Spec} reports the detailed results from the broad-band X-ray spectral fitting while the resulting X-ray light-curve of SN\,2014C is portrayed in Fig. \ref{Fig:XrayBol}. 

Finally, we find evidence an excess of emission around $\sim6.7-6.9$ keV that we identify with H- and He-like transitions in Fe atoms. The Fe emission, as revealed by both the CXO and NuSTAR,  is present in each of the three epochs of observations (Fig. \ref{Fig:ChandraNustar} and \ref{Fig:SED}) with no detectable evolution from one epoch to the other. The results from a spectral line fitting with a Gaussian profile are reported in Table \ref{Tab:Fe}. Our observations do not have the spectral resolution and statistics to resolve what is likely to be a complex of emission lines originating from highly ionized Fe atom states, as suggested by the calculations by \cite{Mewe85}, \cite{Mewe86} and \cite{Liedahl95} (e.g. the MEKAL model within Xspec).
\section{Explosion parameters}
\label{SubSec:ExplPar}

\begin{figure}
\vskip -0.0 true cm
\centering
\includegraphics[scale=0.56]{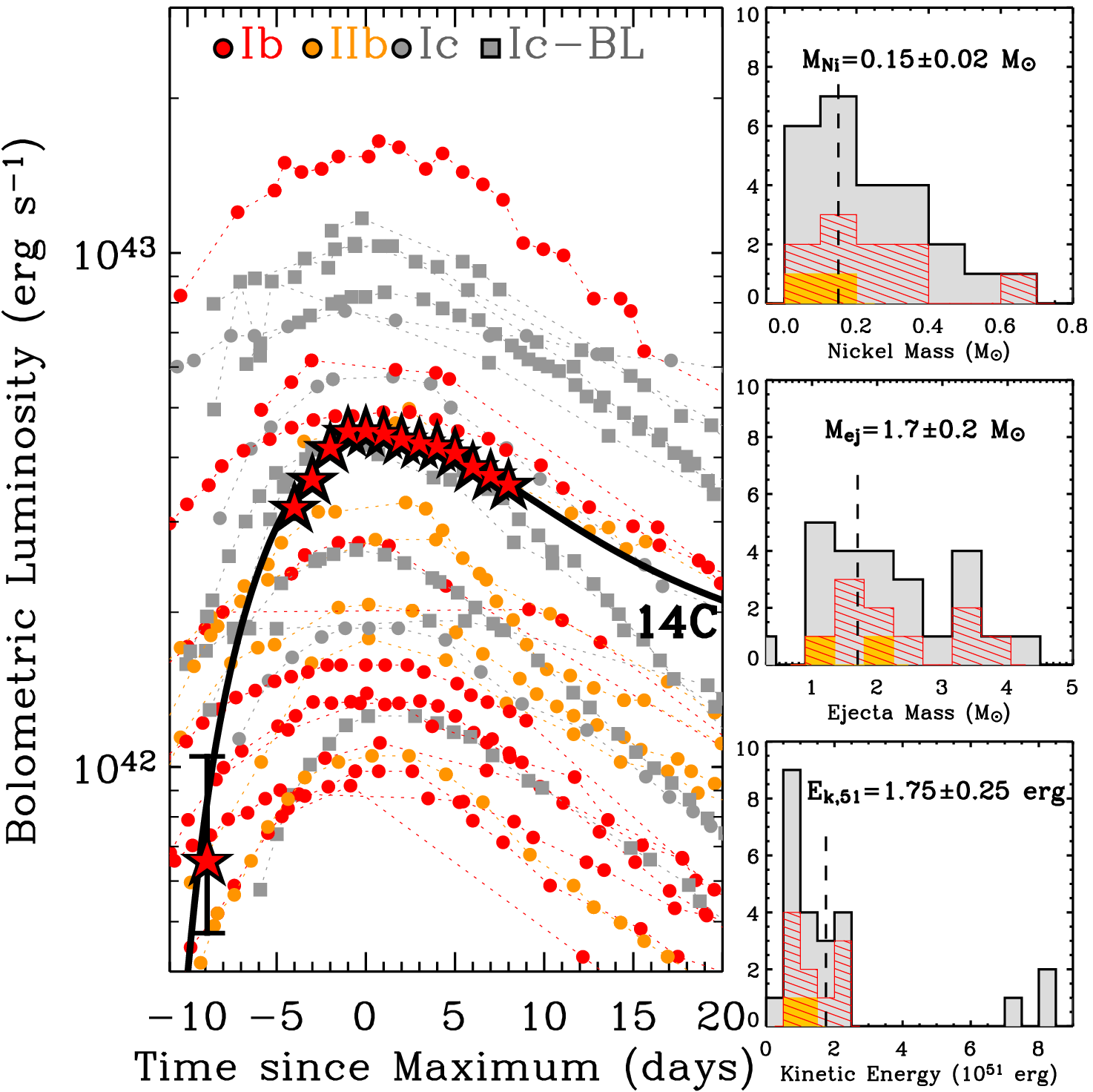}
\caption{\emph{Left Panel:} Bolometric luminosity and best-fitting model (thick black line) of SN\,2014C in the context of well-monitored core-collapse SNe originating from stellar progenitors  that had lost most of their hydrogen envelope before exploding (i.e. type Ic, Ic-BL, Ib and IIb). Data for the other SNe are from \cite{Cano13}, \cite{Cano14}, \cite{Cano14b} and \cite{Lyman14}. \emph{Right Panels:} Explosion parameters of SN\,2014C (vertical dashed lines) compared to the sample of \cite{Drout11}. SN\,2014C shows normal explosion parameters. }
\label{fig:Lbol}
\end{figure}

We calculate the bolometric luminosity of SN\,2014C by integrating the extinction-corrected flux densities in the $v$, $b$ and $u$ UVOT bands and by applying a bolometric correction which corresponds to   effective black-body temperatures in the range $7000-10000$ K. We complement this data set with public photometry, which allow us to constrain the very early light-curve. Specifically, SN\,2014C was first detected on 2.10 UT January, 2014. \cite{Kim14} reports a detection of SN\,2014C at the level of $R=17.1$ mag. Assuming a bolometric correction appropriate for a temperature of emission $T\sim10000-15000$ K, we derive $L_{\rm{bol}}=(0.5-10)\times 10^{41}\,\rm{erg\,s^{-1}}$ at $t\sim-10$ days since maximum light.
Figure \ref{fig:Lbol} shows the resulting bolometric emission from SN\,2014C. With $L_{pk}\sim5\times10^{42}\,\rm{erg\,s^{-1}}$, SN\,2014C shows an intermediate peak luminosity among the sample of well-monitored H-poor SNe of Fig. \ref{fig:Lbol}.

In M15 we showed that before the onset of strong SN shock interaction with a massive CSM around $t\sim100$ days, SN\,2014C exhibited typical spectral features of the class of type Ib SNe (i.e. the end points of the evolution of stellar progenitors that managed to shed their hydrogen envelope, while retaining a helium layer). 
In the absence of strong interaction,  the light-curves of SNe that originate from hydrogen-stripped progenitors are powered by the radioactive decay of $^{56}\rm{Ni}$. Specifically, the optical peak luminosity directly reflects the amount of $^{56}\rm{Ni}$ produced by the explosion ($M_{\rm{Ni}}$), while the light-curve width $\tau$ is sensitive to the photon diffusion time scale and thus to the explosion kinetic energy ($E_{\rm{k}}$) and ejecta mass ($M_{\rm{ej}}$). We employ the analytical model by \cite{Arnett82} with the updated formalism by \cite{Valenti08} and \cite{Clocchiatti97} to estimate the explosion parameters of SN\,2014C. 

The spectra acquired around the time of maximum light indicate a photospheric velocity $v_{\rm{phot}}=13000\,\rm{km\,s^{-1}}$ (M15). We use  $v_{\rm{phot}}$ as characteristic velocity $v_*$ of the ejecta to break the degeneracy between $M_{\rm{ej}}$ and $E_{\rm{k}}$, where\footnote{We replaced the inaccurate numerical factor given in Eq. 65 of \cite{Arnett82} with the correct value, as explained in \cite{Wheeler14}.}  $v_*=(10\,E_{\rm{k}}/3\,M_{\rm{ej}})^{0.5}$ and assume a constant effective opacity $k_{\rm{opt}}=0.07\,\rm{cm^{2}\,g^{-1}}$. Modeling of the bolometric light-curve constrains the time of first light of SN\,2014C to 30 December 2013 $\pm1$ day (MJD 56656 $\pm1$) and yields the following estimates for the explosion parameters: $M_{\rm{Ni}}=0.15\pm0.02\,\rm{M_{\sun}}$, $E_{\rm{k}}=(1.75\pm0.25)\times 10^{51}\,\rm{erg}$, $M_{\rm{ej}}=1.7\pm0.2\,\rm{M_{\sun}}$. The comparison to the sample of type Ib/c SNe in Fig. \ref{fig:Lbol} leads to the conclusion that the explosion parameters of SN\,2014C are typical of the class of SNe with hydrogen-stripped progenitors (\citealt{Drout11}, \citealt{Cano13}, \citealt{Lyman14}). 

As a caveat we note that this analytic treatment is sensitive to
$k_{\rm{opt}}M_{\rm{ej}}$ and $k_{\rm{opt}}E_{\rm{k}}$. As \cite{Wheeler14} showed, a way to solve for this model degeneracy is by using the late-time light-curve decay slope \emph{under the assumption that it is entirely powered by the
radioactive decay of $^{56}\rm{Ni}$ and its products}. This assumption does not hold for SN\,2014C, which is clearly dominated by interaction at late times, as shown in Fig. \ref{Fig:XRO}. For ordinary Ibc SNe \cite{Wheeler14} find $k_{\rm{opt}}$ values as low as $0.02\,\rm{cm^{2}\,g^{-1}}$ (e.g. for SN\,1994I, their Table 2). For SN\,2014C this low value of $k_{\rm{opt}}$ would imply $E_{\rm{k}}\sim10^{52}\,\rm{erg}$. Energetic SNe with $E_{\rm{k}} \sim10^{52}\,\rm{erg}$ are accompanied by broad spectroscopic features that are \emph{not} observed in SN\,2014C (M15). We thus conclude that for SN\,2014C it is likely that $E_{\rm{k}}<10^{52}\,\rm{erg}$ and the effective opacity is $k_{\rm{opt}}>0.02\,\rm{cm^{2}\,g^{-1}}$.

In the following we use 30 December, 2013 as explosion date of SN\,2014C. The possible presence of a ``dark phase''  (e.g. \citealt{Piro12,Piro13}) with duration between hours and a few days between the explosion and the time of the first emitted light has no impact on our conclusions.

\section{Environment}
\label{Sec:Env}

In this section we employ the knowledge of SN\,2014C explosion parameters determined in Sec. \ref{SubSec:ExplPar} and the observables from our broad-band X-ray campaign of Sec. \ref{Sec:obs} to constrain the density profile of the environment around SN\,2014C, shaped by the recent mass-loss history of its progenitor system. The  picture that emerges from our analysis, detailed below, is that of a low-density cavity, with density typical of the environments of ordinary Ibc SNe (e.g. \citealt{Soderberg07}), surrounded by a dense shell, with density comparable to the environments of IIn SNe (e.g. \citealt{Kiewe12}). 

\subsection{Low Density Cavity at $R\lesssim 2\times10^{16}$ cm}
\label{SubSec:IC}
At early epochs ($t\lesssim30$ days), the X-ray emission from SNe originating from H-stripped progenitors is dominated by Inverse Compton (IC) processes (e.g. \citealt{Bjornsson04}). Photospheric optical photons are upscattered to X-ray energies by relativistic electrons accelerated by the SN shock.  IC emission depends on the density structure of the SN ejecta,  the properties of the explosion, and the characteristics of the medium around the SN (e.g. \citealt{Chevalier06}). 

Following \cite{Matzner99}, we assume a SN outer density structure $\rho_{SN}\propto R^{-n}$ with $n\sim9$, as appropriate for stellar explosions arising from compact progenitors. The SN shock accelerates electrons into a power-law distribution $n_e(\gamma)\propto \gamma^{-p}$, where $\gamma$ is the electron Lorentz factor. Well studied SNe Ib/c indicate $p\sim3$, with a fraction of post-shock energy into electrons $\epsilon_e\sim 0.1$ (e.g. \citealt{Chevalier06}). Our modeling of Sec. \ref{SubSec:ExplPar} constrains the explosion kinetic energy $E_{\rm{k}}=(1.75\pm0.25)\times 10^{51}\,\rm{erg}$ and ejecta mass $M_{\rm{ej}}=1.7\pm0.2\,\rm{M_{\sun}}$, which values we employ here.  Finally, the last stages of evolution of massive stars are predicted to be characterized by powerful winds which are expected to shape the immediate SN environment  within $R\sim4\times 10^{16}\,\rm{cm}$ (e.g. \citealt{Ramirez01}, \citealt{Dwarkadas07}) into a density profile $\rho_{CSM}\propto R^{-2}$.

By employing the IC formalism from \cite{Margutti12} and the optical bolometric emission from SN\,2014C of Sec. \ref{SubSec:ExplPar}, we find that the lack of detectable X-ray emission from SN\,2014C during the first $\sim20$ days (Sec. \ref{SubSec:XRT}) implies a low density environment at distances $R\sim (0.8-2)\times 10^{16}\,\rm{cm}$. The inferred mass-loss rate is $\dot M<(3-7)\times 10^{-6}\,\rm{M_{\sun}\,yr^{-1}}$ for an assumed wind velocity $v_w=1000\,\rm{km\,s^{-1}}$. The finding of a low density environment in the proximity of SN\,2014C directly constrains the mass-loss history of its progenitor star: in particular, it implies that the progenitor did not suffer massive eruptions within $\Delta t=7(v_w/1000\,\rm{km\,s^{-1}})\,\rm{yrs}$ before the final explosion. 


\subsection{Region of H-rich Material with Enhanced Density at $R\sim5.5\times 10^{16}$ cm}

The rising X-ray and radio luminosity of the type Ib SN\,2014C at late times, coupled with the progressive emergence of prominent H$\alpha$ emission (Fig. \ref{Fig:XRO}),  clearly suggests a scenario where the freely expanding, H-poor SN\,2014C ejecta encountered a dense H-rich region  in the proximity of the explosion site. We constrain the properties of the H-rich, dense CSM shell by using the following observables:
\begin{itemize}
\item Optical spectroscopy that we presented in M15 constrains the emergence of H$\alpha$ emission due to the interaction of the SN ejecta with H-rich material in the CSM to $t>30$ days. A prominent H$\alpha$ profile has developed by day $130$ after the explosion. In the following we use $t=130$ days as the start time of the \emph{strong} CSM interaction. 
\item The broad-band X-ray luminosity shows a sharp rise during the first  $\sim300$ days and reaches its maximum value $L_x\sim5\times 10^{40}\,\rm{erg\,s^{-1}}$ by $\sim$500 days.
\item The X-ray emission is of thermal origin and the observed temperature is $T\sim 20$ keV between 400 and 500 days after the explosion.
\item There is significant evidence for decreasing absorption with time, with $\rm{NH_{tot}}$ evolving from  $\sim 4\times 10^{22}\,\rm{cm^{2}}$ at $\sim300$ days, to $\sim 2\times 10^{22}\,\rm{cm^{2}}$ at $\sim500$ days after the explosion.
\end{itemize}

The interaction of freely expanding SN ejecta with the CSM leads to the formation of a double shock interface layer, with the forward shock (FS) propagating into the CSM and the reverse shock (RS) decelerating the SN ejecta. For the SN ejecta we adopt a power-law density distribution with  $\rho_{SN}=\rho_0(v/v_0)^{-n}$ for $v>v_0$ and $\rho_{SN}=\rho_0$ for ejecta velocities $v<v_0$. For the explosion parameters of SN\,2014C 
the transitional velocity is $v_0\sim10700\,\,\rm{km\,s^{-1}}$. We assume $n=9$ as appropriate for compact stellar progenitors \citep{Matzner99}. We follow \cite{Chevalier82}  to describe the dynamics of the shock propagation into the low density bubble, and \cite{Chevalier89}  to compute the dynamics of the strong interaction of the SN ejecta with the dense shell. 
\subsubsection{Expansion in the Bubble}
\label{SubSubSec:bubble}
The dynamics of the double shock structure that originates from the interaction of the outer power-law portion of the SN ejecta profile with a wind-like CSM with density $\rho_{CSM}=\dot M/(4\pi v_w R^{2}$) is described by a self-similar solution \citep{Chevalier82}. From Sec. \ref{SubSec:IC},  $\dot M<(3-7)\times 10^{-6}\,\rm{M_{\sun}\,yr^{-1}}$ for $v_w=1000\,\rm{km\,s^{-1}}$. With these parameters, we find that the FS radius at the time of the start of the strong interaction ($t=130$ days) is $\sim5.5\times 10^{16}\,\rm{cm}$. The swept up mass of gas within the bubble is $M_{\rm{bubble}}<10^{-4}\,\rm{M_{\odot}}$. $M_{\rm{bubble}}$ is considerably smaller than the mass in the outer power-law section of the SN ejecta ($M_{ej,PL}\sim0.6\,\rm{M_{\odot}}$), which is thus only minimally decelerated during its expansion into the cavity. The velocity of the FS just before the start of the strong interaction is $\sim44000\,\rm{km\,s^{-1}}$. The CSM density at the outer edge of the bubble is $<100\,\rm{cm^{-3}}$.

\subsubsection{Interaction with the dense, H-rich CSM}
The rising X-ray and radio luminosity with time is suggestive of a large density contrast between the bubble and the shell of dense CSM at $R_{shell}\sim5.5\times 10^{16}\,\rm{cm}$, as confirmed by our modeling below. Under these circumstances (i.e. $\rho_{\rm{bubble}}\ll\rho_{\rm{shell}}$ and $M_{ej,PL}\gg M_{bubble}$), the outer density profile of the SN ejecta continues to interact with the dense shell of CSM and the energy transmitted to the shell is initially modest \citep{Chevalier89}. The analysis by \cite{Chevalier89} shows that the flow once again has a self-similar nature, with $R_{RS}=0.92\,R_{shell}$ for $n=9$ (their Table 2). 

The collision of the SN ejecta with the dense CSM shell causes a sudden increase of the X-ray luminosity of both the RS and the FS (see e.g. \citealt{Chugai06}, their Fig. 2). The FS experiences rapid deceleration, with $v_{FS}$ decreasing from $\sim44000\,\rm{km\,s^{-1}}$ to a $\sim$ few $1000\,\rm{km\,s^{-1}}$, as suggested by the measured width of the intermediate component of the H$\alpha$ line in our spectra, which maps the dynamics of the shocked, H-rich CSM shell (Fig. \ref{Fig:XRO}, M15). Since the velocity of the shock determines the energy imparted to particles that cross the shock, the characteristic temperature of emission of material in the FS plummets: contrary to the previous phase of expansion within the bubble, the interaction with a wall of material causes the FS temperature to be significantly below the temperature of the RS. During the expansion in a wind-like medium $T_{RS}/T_{FS}\propto v_{RS}^2/v_{FS}^2$  or $T_{RS}\propto T_{FS}/(n-3)^2\sim 0.03 T_{FS}$ for $n=9$, see e.g. \cite{Chevalier03}.

This rapid deceleration is followed by a period of steady acceleration as the faster SN ejecta piles up from behind (\citealt{Chevalier89}, \citealt{Dwarkadas05}). The radius of the shock wave expands as $R_{FS}=R_{shell}+K_1(A\,R_{shell}^{2-n}t^{n-3}/\rho_{shell})^{1/2}$, where $K_1=0.26$ for a nonradiative shock and $K_1=0.24$ for a radiative shock front, $n=9$, $A\equiv A(E_{\rm{k}},M_{\rm{ej}},n)$  and we have assumed a smooth CSM shell with density $\rho_{shell}$ (\citealt{Chevalier84}, \citealt{Chevalier89}).

We proceed within the ``thin shell'' approximation (e.g. \citealt{Chevalier82}), where the shocked gas, made by shocked CSM \emph{and} shocked SN ejecta, can be treated as a thin shell with mass $M_s$, radius $R_s (t)$ and velocity $v_s(t)$.  $v_{FS}\approx v_s$, the velocity of the (unshocked) ejecta at the RS is $v_{SN}=R_s/t$ and $v_{RS}\approx v_{SN}-v_{FS}$. For a decelerated FS velocity $v_{FS}\approx $ a few $1000\,\rm{km\,s^{-1}}$, we estimate at $t\sim 400$ days  $R_s\sim6\times 10^{16}\,\rm{cm}$ and $v_{RS}\sim13000\,\rm{km\,s^{-1}}$. The RS temperature is directly connected to its velocity by $T_{RS}=2.27\times 10^9 \mu_p v_{RS,4}^2$, where $v_{RS,4}\equiv v_{RS}/10^4\,\rm{km\,s^{-1}}$ and $\mu_p$ is the mean mass per particle including electrons and ions (e.g. \citealt{Fransson96}). We follow \cite{Chugai06} and approximate the shocked SN ejecta by a mixture of  65\% He, 33\% O and 2\% Fe by mass. For this composition of the ejecta and complete ionization, $\mu_p=1.33$ and  $T_{RS}>400\,\rm{keV}$, which is much larger than the $T\sim 18$ keV indicated by our broad-band X-ray spectral analysis (incomplete ionization would lead to even larger temperatures).  This finding thus suggests that the detected X-ray emission is dominated by the FS. Under this hypothesis, $T_{FS}=18$ keV at $t\sim400$ days, which implies $v_{FS}\sim4000\,\rm{km\,s^{-1}}$, consistent with the  indication of $v_{FS}\sim$ few $1000\,\rm{km\,s^{-1}}$ from the optical spectra. Solar abundances have been assumed for the CSM (i.e. $\mu_p=0.61$).

The mass of the shocked CSM material is directly constrained by the observed bremsstrahlung spectrum. The observed X-rays at $t\sim500$ days, with $T\sim 20$ keV and $L_x\sim5\times 10^{40}\,\rm{erg\,s^{-1}}$ require an emission measure $EM_{FS}\sim1.4\times 10^{63}\,\rm{cm^{-3}}$, where $EM\equiv \int n_e n_I dV$. Accounting for the presence of an additional thermal component of emission from the RS at $T\gg20$ keV reduces the required EM to $EM_{FS}\sim 1.1\times 10^{63}\,\rm{cm^{-3}}$ and suggests $EM_{RS}\sim 4.3\times 10^{62}\,\rm{cm^{-3}}$. To estimate the mass of the shocked CSM gas we need to constrain the volume of the shocked CSM. Interpreting the occurrence of the peak of the emission at $t\sim500$ days as due to the passage of the shock front through the CS shell, we constrain the shell thickness $\Delta R_{shell}\sim10^{16}\,\rm{cm}$.  The mass of the shocked CSM is thus $M_{CSM}\sim(1.0-1.5)\,\rm{M_{\odot}}$ with density $\rho_{shell}\sim2\times 10^6\,\rm{cm^{-3}}$. Introducing a volume filling factor $f$ defined as $V_{shell}=4\pi R_{shell}^2 \Delta R_{shell} f$, the previous estimates would scale as:$M_{CSM}\propto f^{1/2}$ and $\rho_{shell}\propto f^{-1/2}$.

The volume of the reverse postshock layer between the reverse shock and the contact surface can be easily computed considering that for $n=9$, $R_{RS}=0.92\,R_{shell}$ \citep{Chevalier89}. For $EM_{RS}\sim 4.3\times 10^{62}\,\rm{cm^{-3}}$ the mass of the shocked SN ejecta is thus $M_{ej,RS}\sim0.7\,\rm{M_{\odot}}$. As a sanity check we note that momentum conservation implies that the mass of the CSM required to decelerate $\sim0.7\,\rm{M_{\odot}}$ of SN ejecta with typical velocity $(2E_{\rm{k}}/M_{\rm{ej}})\gtrsim 10^4\,\rm{km\,s^{-1}}$ down to $\sim4000\,\rm{km\,s^{-1}}$ is $M_{shell}\gtrsim 1.4 M_{ej,RS}$ or $M_{shell}\gtrsim 1\,\rm{M_{\odot}}$, consistent with our estimates above.

From another perspective, since the observed emission is dominated by the FS, the detected decrease of X-ray absorption with time  (Fig. \ref{Fig:ChandraNustar}) provides an independent constraint to  the amount of neutral CSM material in front of the FS.\footnote{Note that the total amount of material is likely larger, as some material will be ionized.} The detected temporal variation of $\rm{NH_{tot}}$ directly implies that the material responsible for the absorption is local to the SN explosion and within the reach of the SN shock over the time scale of our observations. From $t=308$ days to $472$ days after the explosion we measure $\Delta \rm{NH_{tot}}\sim 2\times 10^{22}\,\rm{cm^{-2}}$. For $v_{FS}\sim4000\,\rm{km\,s^{-1}}$, the detected $\Delta \rm{NH_{tot}}$ constrains the neutral CSM mass probed by the shock front between 308 days and 472 days to be $M_{CSM,NH}\sim0.6\,\rm{M_{\odot}}$, while the total CSM shell mass is $\gtrsim1.2\,\rm{M_{\odot}}$, assuming that the FS did not experience substantial acceleration and the CSM shell is spherical and homogeneous. 

Finally, while a detailed hydrodynamical simulation is beyond the scope of the present work, we end the section emphasizing the qualitative agreement of our conclusions, derived from a purely analytical treatment,  with the results from the simulations from \cite{Chugai06}. In order to reproduce the properties of SN\,2001em, \cite{Chugai06} simulated the collision of freely expanding SN Ibc ejecta with a dense shell of H-rich material at $R_{shell}=(5-6)\times 10^{16}\,\rm{cm}$ with thickness $\Delta R_{shell}\sim 10^{16}\,\rm{cm}$ and $M_{shell}=(2-3)\,\rm{M_{\odot}}$. The simulation thus differs from our situation only in terms of the larger mass of the CSM shell. These authors find that the SN ejecta interaction with the dense medium causes a large increase of $L_x$ of both shocks, with $L_x$ reaching $L_x\sim 10^{41}\,\rm{erg\,s^{-1}}$ at peak.  The FS, which was hotter than the RS before the strong interaction, experiences rapid deceleration, followed by  a period of acceleration until the shock front reaches the edge of the shell. As a result, $T_{FS}\ll T_{RS}$ after the interaction (e.g. at $t=1000$ days, $T_{FS}\sim 5$ keV and $T_{RS}\sim 100$ keV, and $T_{RS}\sim850$ keV at $t=500$ days, see their Fig. 2). Compared to SN\,2014C, the FS in the simulations by \cite{Chugai06} is more strongly decelerated by the impact with the CSM shell, due to the larger mass of the shell (the SN ejecta parameters are instead comparable). For the same reason, in their simulations  the peak of the X-ray emission due to the passage of the shock front through the CSM shell is also delayed with respect to SN\,2014C ($\sim1000$ days, vs. $\sim500$ days).  Apart from these \emph{expected} differences,  our analytical treatment captures the key physical properties of the SN ejecta- CSM strong interaction. 

\subsubsection{Anticipated Evolution at Later Times}
\label{SubSubSec:Later}
The shock acceleration phase, caused by the increase of the pressure in the shocked region due to the interaction with the outer density profile of the SN ejecta, ends when (i) the RS reaches the flat portion of the ejecta profile; or (ii) the energy transmitted to the CSM shell becomes large compared to the energy of the shocked ejecta; or, alternatively, (iii) the shock front reaches the edge of the CSM shell  \citep{Chevalier89}.

The time scale at which the RS reaches the bend in the SN ejecta profile is $t_1\sim R_{RS}/v_0$, where $v_0\equiv v_{0}(E_{\rm{k}},M_{\rm{ej}},n)$ is the transitional velocity that defines the SN ejecta profile and $v_0\sim10700\,\rm{km\,s^{-1}}$ for the explosion parameters of SN\,2014C. Following \cite{Chevalier89}, $t_1\sim (R_{shell}/v_0)(1+ 3\gamma (\gamma-1)/((\gamma+1)(n-5)))^{-1/3}\approx0.92(R_{shell}/v_0)$ for $n=9$ and an adiabatic index $\gamma=5/3$. For SN\,2014C we thus derive $t_1\gtrsim 550$ days.

The time scale $t_2$ at which the energy transferred to the CSM gas is $\ge0.5$ the total energy in the shocked region depends on the explosion parameters and on the properties of the CSM shell $t_2\equiv t_2(M_{\rm{ej}},E_{\rm{k}},n,\rho_{shell}, R_{shell})$. Employing the formalism by \cite{Chevalier89}, their Eq. 3.24, we find $t_2>550$ days for $\rho_{shell}\ge 4\times 10^6\,\rm{cm^{-3}}$.

$L_x$ reaches its maximum at $t<500$ days (Fig. \ref{Fig:XrayBol}), from which we deduce that the shock front already reached the edge of the high-density region by $t_3\sim500$ days. The later dynamics of the interaction is thus likely determined by this event, even if the time scales $t_1$ and $t_2$ are close enough that a simulation is needed to capture the details of the evolution.

Once the shock has transversed the dense shell, a rarefaction wave propagates into the interaction region, while the FS expands into a less dense medium that was shaped by the previous phase of evolution of the progenitor of SN\,2014C. The simulations by \cite{Chugai06} show that the X-ray luminosity remains high even after the shell has been overtaken, but that the temperature of emission of both shocks declines with time. According to our calculations, for SN\,2014C, the temperature of the RS shock will enter the NuSTAR passband at  $t>800$ days ($T_{RS}\sim100$ keV at $t\sim800-1000$ days, depending on the ionization state of the ejecta). Future observations will allow us to sample the density of the CSM outside the dense shell. At the moment we note that  optical spectroscopy of SN\,2014C (M15) reveals that the material outside the CSM shell is H-rich and shows velocities $<100\,\rm{km\,s^{-1}}$ (from the narrow component of the H$\alpha$ line, Fig. \ref{Fig:XRO}). These velocities are typical of winds emanating from non-compact progenitors, and are typically associated with the large mass-loss rates (and densities) inferred for type IIn SNe (e.g. \citealt{Kiewe12}). It is thus possible that the shocked gas will never re-enter a phase of free expansion. 
\subsubsection{Clumpy Structure of the CSM}
Two independent sets of observations point to a complex structure of the interaction region with overdense clumps of emitting material: (i) the velocity profile of the shocked H-rich material; (ii) the presence of prominent Fe emission lines in the X-ray spectra.

The progressive emergence of an H$\alpha$ emission (Fig. \ref{Fig:XRO}) indicates the presence of shocked H-rich CSM. However, optical spectroscopy in Fig. \ref{Fig:XRO} (see M15 for details) shows velocities  $<2000\,\rm{km\,s^{-1}}$ for the shocked material, while above we infer  $v_{FS}\sim4000\,\rm{km\,s^{-1}}$ for the expansion of the shock front at the same epoch. These two results can be reconciled if the H-rich material that dominates the H$\alpha$ emission is clumpy and is concentrated in regions with density contrast $\rho_{clumps}/\rho_{shell}\approx 4$.

Astrophysical plasmas with $T<3\times10^7$ K produce a forest of X-ray lines.  At higher temperatures line emission is inhibited and cooling is dominated by bremsstrahlung. At $T\sim 20 $ keV  (i.e. $\sim2\times 10^8$ K) the only transitions that survive are those associated with extremely ionized states of Fe atoms, i.e. He-like and H-like Fe atoms. Consistent with these expectations, in each epoch of observation we clearly identify a localized excess of X-ray emission at $\sim6.7-6.9$ keV that we associate with He-like and H-like Fe atoms transitions (Fig. \ref{Fig:ChandraNustar} and \ref{Fig:SED}). However, a single-temperature, collisionally-ionized plasma model in thermal equilibrium fails to reproduce the observed luminosity of the Fe emission at all epochs. At temperature $\sim20$ keV and density $\sim4\times10^6\,\rm{cm^{-3}}$ the time for electrons and protons to come into equilibrium is $t_{ei}<t_{obs}$ and $t_{ei}\sim80$ days \citep{Spitzer62},  justifying our assumption of thermal equilibrium so far. The only way to reconcile the prominent Fe emission within this model is to invoke extremely super solar abundances  ($\sim5$ times the solar value) for the shocked CSM shell. Alternatively, observations are consistent with a multi-phase plasma, with a lower temperature (and higher density) component responsible for the Fe emission. Given the independent suggestion of a clumpy medium from the H$\alpha$ velocity profile, we consider this second possibility of a medium with components of different densities (and not necessarily spherically simmetric) more likely.

Evidence or hints for an excess of emission around $6.7-6.9$ keV has been found in some SNe characterized by strong interaction with a dense medium. Examples include the type-IIn SNe 1996cr (\citealt{Dwarkadas10}, \citealt{Dewey11}), 2006jd \citep{Chandra12b} and SN\,2009ip \citep{Margutti14}.  In all cases the excess has been interpreted as originating from H- and He-like transitions of collisionally ionized Fe atoms. Interestingly, in the case of SN\,2006jd, \citealt{Chandra12b} arrived to a similar conclusion of either a very enriched medium with super solar abundances or a multi-phase plasma to explain the luminous Fe emission.

\section{SN2014C in the context of 183 Ib/c SNe with radio observations}
\label{Sec:Rates}

\begin{figure*}
\vskip -0.0 true cm
\centering
\includegraphics[scale=0.65]{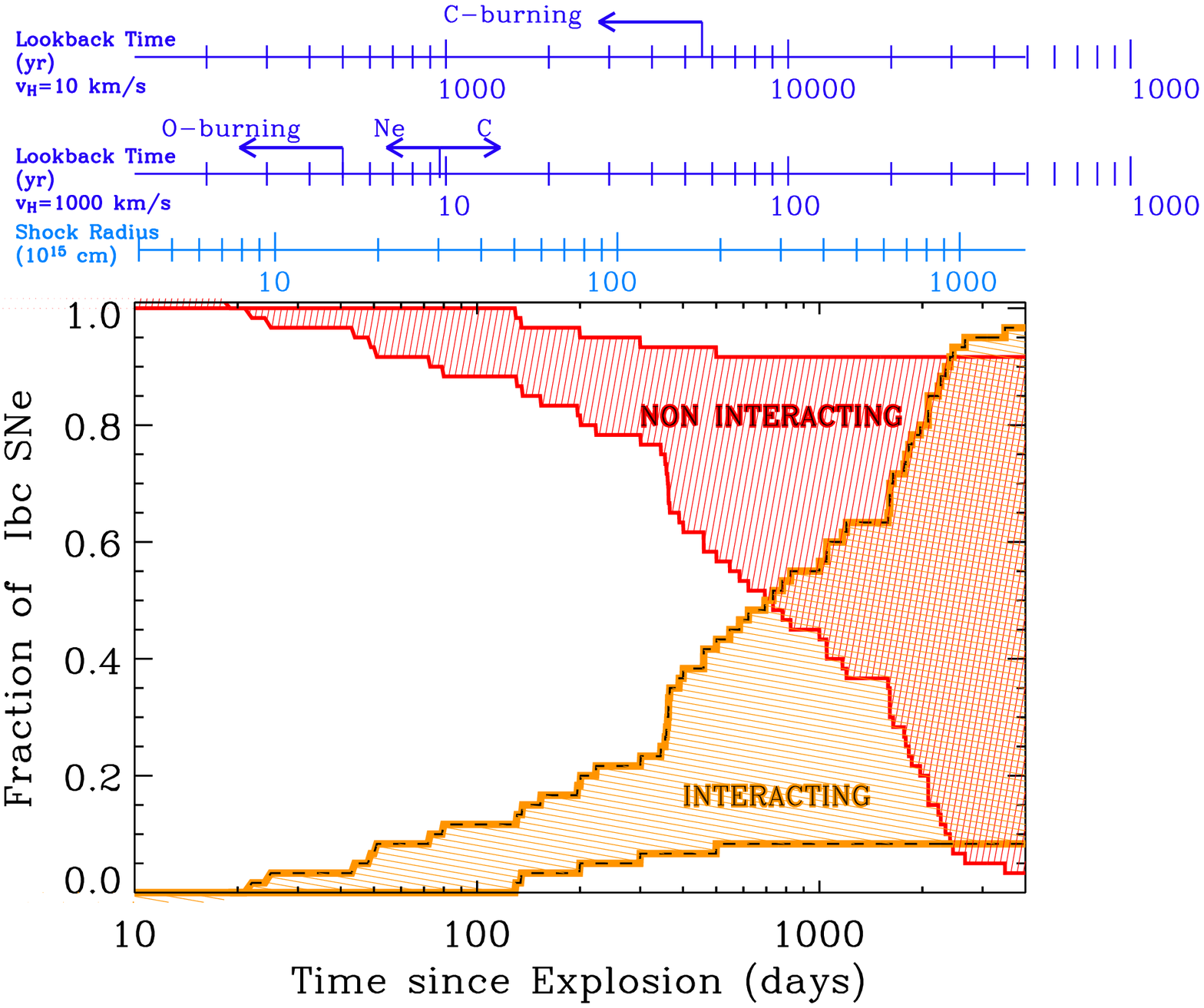}
\caption{Constraints to the fraction of Ib/c SNe that are interacting (orange shaded area) or non interacting (red shaded area) with a14C-like medium as a function of time since the explosion, as derived from the analysis of 60 SNe of type Ib/c with constraining radio observations. For this sample, the fraction of objects that does not show signs of interaction at very early times is 100\% by definition, as we selected spectroscopically classified type Ib/c SNe (SNe with signs of interaction since the very first moment would be instead classified as IIn or Ibn events depending on the H-rich or He-rich composition of the medium, respectively) The time since the explosion is converted into a shock radius by employing a standard shock velocity of $0.15c$. We show the lookback time for two representative ejection velocities of the H-rich material $v_{\rm{H}}$. The corresponding nuclear burning stages are for a non-rotating stellar progenitor of $12\,\rm{M_{\odot}}$ with solar metallicity from \cite{Shiode14}.
}
\label{Fig:Rates}
\end{figure*}

\begin{figure}
\vskip 0.2 true cm
\centering
\includegraphics[scale=0.5]{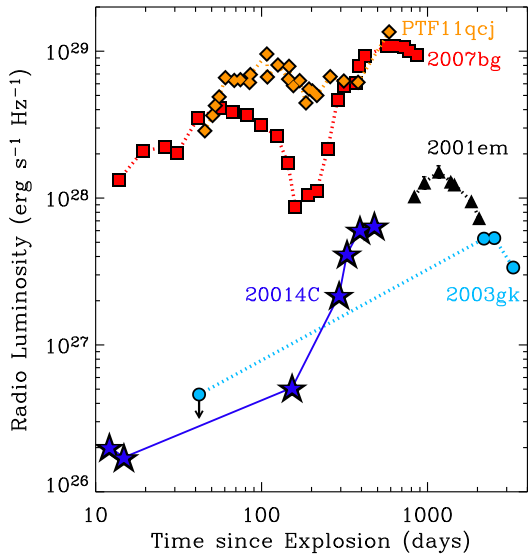}
\caption{The type Ib/c SNe 2001em, 2003gk, 2007bg and PTF11qcj display late-time radio re-brightenings with similarities to SN\,2014C. 8.5 GHz data have been shown for SNe 2001em, 2003gk and 2007bg (\citealt{Schinzel09}, \citealt{Bietenholz14}, \citealt{Salas13} ). For PTF11qcj we show here the 7.4 GHz data from \cite{Corsi14}. For SN\,2014C we use observations acquired at 7.1 GHz (K15).
}
\label{Fig:14Clike}
\end{figure}

\subsection{Rate of 14C-like explosions among Ib/c SNe}
\label{SubSec:rates14C}

The metamorphosis of SN\,2014C from an ordinary type-Ib SN into a strongly interacting SN of type IIn over a time scale of $\sim1$ yr is signaled by the progressive emergence of H$\alpha$ emission and narrow line emission in the optical/UV band, accompanied by a  marked increase of both the radio and X-ray emission produced by the SN shock interaction with a ``wall'' of dense H-rich material presumably deposited in the environment by the stellar progenitor before death (Fig. \ref{Fig:XRO}). In this section we quantify how common this phenomenology is among SNe originating from H-stripped stars by using the available set of published data from radio monitoring campaigns of type Ib/c SNe. We focus on the radio wavelength range as it offers the most homogeneous data set with observations covering the early- and late-time evolution of Ib/c SNe.

Our sample comprises public observations of 183 type Ib/c SNe obtained over more than 20 years of radio monitoring campaigns, with data acquired from a few days to $\sim32$ yrs since the SN explosion. The data have been collected from \cite{Berger03a}, \cite{Soderberg06b}, \cite{Soderberg07}, \cite{Bietenholz14},  \cite{Corsi14}, \cite{Kamble14},  \cite{Drout15} and references therein.   In 41 cases the radio observations provide meaningful constraints to the presence of a 14C-like radio re-brightening (i.e. the observations are deep enough and cover the late-time evolution of the transient at $t\gtrsim500$ days). For the remaining 142 SNe Ib/c  the available observations do not typically reach the necessary depth and/or do not extend to late times. 

Out of 41 type Ib/c SNe with good radio coverage, we can rule out a 14C-like radio re-brightening in 37 cases. For the type Ib/c SNe 2001em (\citealt{Bietenholz05}, \citealt{Schinzel09}), 2003gk (\citealt{Bietenholz14}), 2007bg (\citealt{Soderberg07}, \citealt{Salas13}) and  PTF11qcj (\citealt{Corsi14}) we find evidence for late-time, luminous radio re-brightenings consistent with a 14C-like phenomenology. In particular we note that (i) albeit sparsely sampled, the X-ray and optical evolution of the type-Ic SN\,2001em \citep{Filippenko01} are also reminiscent of SN\,2014C (\citealt{Chugai06}); (ii) we further point out the possible detection of an outburst from the stellar progenitor of PTF11qcj $\sim2.5$ yrs before the explosion (\citealt{Corsi14}).  We thus conclude that $\sim10$\%  of Ib/c SNe with constraining radio observations displays late-time radio re-brigthenings reminiscent of SN\,2014C and that, when available, multi-wavelength observations of this subset of SNe independently support the idea of an enhanced mass loss from the progenitor star during the last stages of evolution preceding core collapse. We note that  while this sample has been collected from different sources, there is no obvious observational bias that would favor a larger fraction of interacting systems. In fact, radio SNe tend to be followed up at later times in the case of an early time radio detection, which suggests that we might have missed later time radio re-brightenings in systems with faint or undetected early emission. This source of bias is alleviated in part by the fact that some SNe in our sample have been followed up at early and late times (irrespective from a detection of lack thereof)  as part of the searches for off-axis emission from a Gamma-Ray Burst-like jet.  We quantify this source of uncertainty in the next paragraph.\footnote{We also would like to mention the existence of SNe (e.g. SNe 1986J and 1996cr, \citealt{Rupen87}, \citealt{Dwarkadas10}) that have been discovered at late times from their bright radio emission and classified as type IIn events from their late-time spectra. These events are not part of this sample by definition. However, we note that they might have been classified as type Ib/c SNe if early optical spectra had been available.}

Figure \ref{Fig:Rates} summarizes the results from the analysis of the enlarged sample of 60  Ib/c SNe for which constraining radio observations have been obtained at some time since the explosion (i.e. we relaxed the condition of a radio monitoring extending to late times $t\gtrsim500$ days of the previous paragraph). This plot shows that existing radio observations rule out the presence of strong interaction in a large fraction of Ib/c SNe \emph{only} at early times (e.g. for $t\lesssim 100$ days, $\gtrsim80$\% of Ib/c SNe do not show evidence for strong shock interaction), while at later times the phase space is sparsely sampled, so that, for example, we can exclude a 14C-like behavior at $t\gtrsim 1000$ days only for $\sim40\%$ of the SNe.  

Finally we note that  SN\,2014C represents an extreme case of radio flux variability. Small-scale radio light-curve modulations at the level of a factor $\sim$2 in flux are common and found in $\sim50$\% of SNe Ib/c with radio detection (see also \citealt{Soderberg07}). This phenomenology can be explained within the context of turbulence-driven small-scale clumping of the stellar wind (\citealt{Moffat08}), a physical process that generates moderate density variations of a factor $\sim 2-4$ in the stellar environment. The ``bubble plus thick shell structure'' that we infer for SN\,2014C clearly demands a different origin. More pronounced achromatic radio flux variations due to modulations of the environment density of a factor $\sim3-6$ have been observed in SNe 2004cc, 2004dk and 2004gq \citep{Wellons12}. In particular, the radio light-curve of SN\,2004cc shows a  well-defined double-peaked structure with flux contrast $\sim10$, a first peak of emission at $\sim25$ days and a second radio peak at $\sim150$ days since explosion (\citealt{Wellons12}, their Fig. 1). While this behavior is somewhat reminiscent of SN\,2014C, the radio flux from SN\,2004cc rapidly and significantly faded on a timescale $\Delta t/t \lesssim 1$, pointing to a smaller mass of the dense region encountered by the SN shock. This comparison highlights the fact that, among type Ib/c SNe, 14C-like events might represent the most extreme manifestations of a more common physical process that induces severe progenitor mass loss synchronized with the final explosion on a variety of mass-loss scales. With this notion in mind, in the following we concentrate on the nature of 14C-like SNe.

\subsection{Statistical inference on the nature of the underlying physical process}
\label{SubSec:Stat}

Current radio studies efficiently sample the first $t\sim 500$ days of evolution of Ib/c SNe (Fig. \ref{Fig:Rates}). The direct implication is that for a typical SN shock velocity of $\sim0.15\,c$  we are currently systematically exploring a region of $\sim2\times10^{17}$ cm around Ib/c SNe. For a medium that has been enriched by material ejected by the stellar progenitor with velocity $v_w$, this fact implies that we are effectively sampling $\Delta t_{sampled}\sim60\times(v_w/1000\,\rm{km\,s^{-1}})^{-1}$ yrs of life of the massive progenitor star before the explosion. This  value corresponds to a very small fraction $f<2\times 10^{-3}$ of the entire life span $\tau$ of a massive star,  even in the case of slowly moving material with $v_w=10\,\rm{km\,s^{-1}}$ and a very massive progenitor with a short life of $\tau\sim 3$ Myrs. We consider $v_w=10-1000 \,\rm{km\,s^{-1}}$ a representative range of velocities of the ejected material.  Velocities of the order of $v_w\sim10\,\rm{km\,s^{-1}}$ are expected in the case of common envelope ejection by a binary system, while $v_w\sim1000\,\rm{km\,s^{-1}}$ are the typical wind velocities observed in Wolf-Rayet stars.


From Sec. \ref{SubSec:rates14C}, the fraction of 14C-like objects is $\gg f$, which implies that 14C-like mass ejections preferentially occur towards the end of the life of the stellar progenitor. We will refer to the interval of stellar lifetime during which a mass ejection can occur as the progenitor ``active time'' ($\Delta t_{active}$). The physical process responsible for the mass ejection is thus closely synchronized with the stellar death.

The results from Sec. \ref{SubSec:rates14C} are statistically consistent with two scenarios. (i) We are sampling a representative portion of $\Delta t_{active}$ of Ib/c SN progenitors and $\Delta t_{active}\sim \Delta t_{sampled}$. Mass ejections are intrinsically rare and only happen in a limited fraction ($\sim10$\%) of progenitors under peculiar circumstances. (ii) $\Delta t_{active}\gg\Delta t_{sampled}$, all Ib/c SN progenitor stars experience an active phase and the small sample of Ib/c SNe with evidence for strong interaction is a mere consequence of our incomplete sampling of the previous stages of stellar evolution. If this is the case  $\Delta t_{active}\sim500\times(v_w/1000\,\rm{km\,s^{-1}})$ yrs. The minimum and maximum $\Delta t_{active}$ consistent with the detection of 4 strong radio re-brigthenings out of a sample of 41 Ib/c SNe can be easily derived from a binomial distribution with $p=\Delta t_{sampled}/\Delta t_{active}$, where $p$ is the probability of success (e.g. \citealt{Romano14}).  For $v_w=1000\,\rm{km\,s^{-1}}$ we find $200\,\rm{yr}<\Delta t_{active}<5000\,\rm{yr}$ (3$\,\sigma$ confidence level), while for $v=10\,\rm{km\,s^{-1}}$  we find $22000 \,\rm{yr}<\Delta t_{active}< 500000\,\rm{yr}$.

We consider hypothesis (ii) the most likely scenario, since we are only sampling a limited portion of the parameter space and we have no reasons to believe that the portion that we are sampling is truly representative of the entire distribution.

\section{Interpretation and Discussion: Massive Star Evolution Revised}
\label{Sec:int}

\begin{figure*}
\vskip -0.0 true cm
\centering
\includegraphics[scale=0.8]{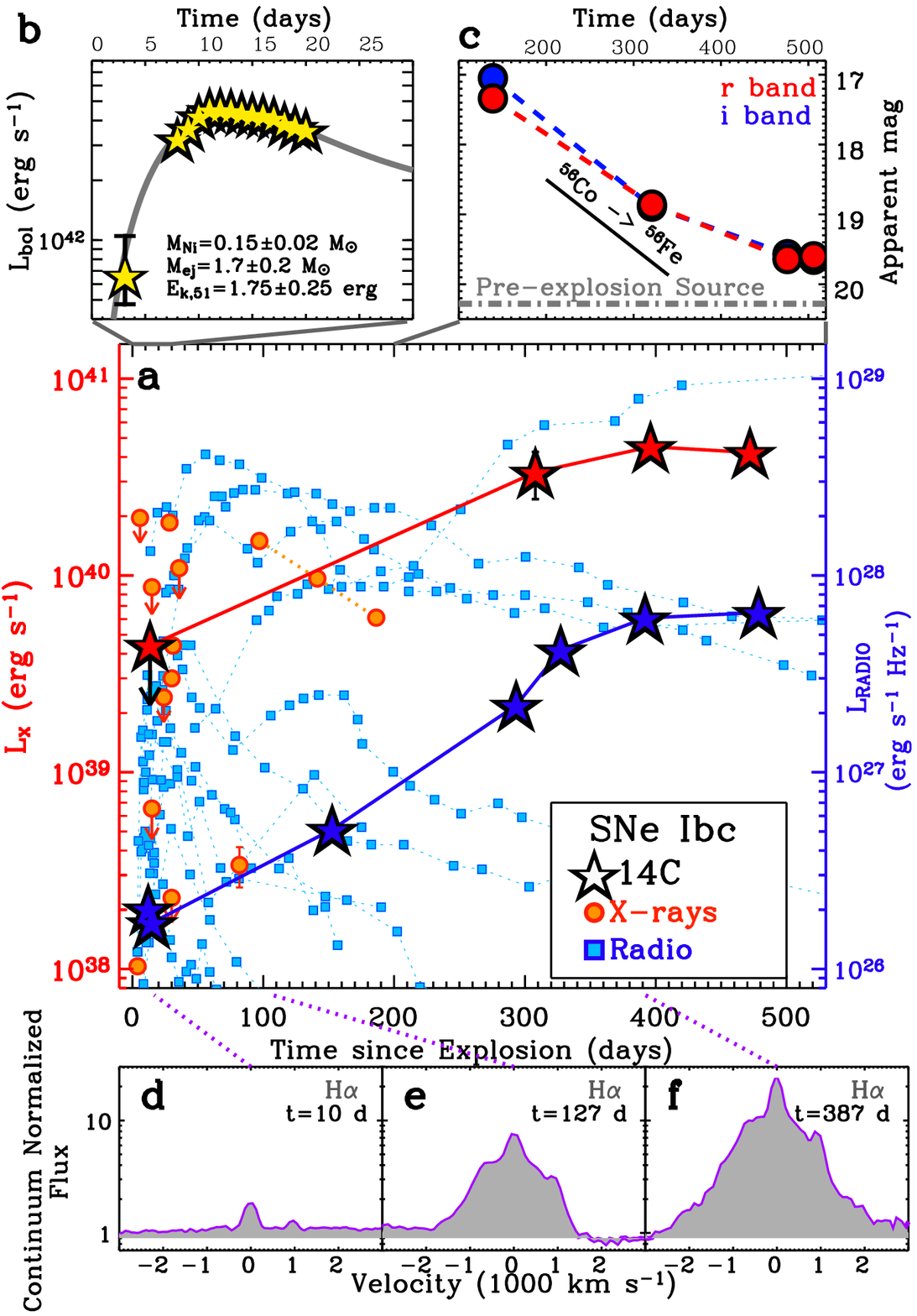}
\caption{This plot summarizes the key and unique observational features of SN\,2014C over the electromagnetic spectrum. \emph{Central Panel:} X-ray (red stars) and radio (7.1 GHz, blue stars) evolution of SN\,2014C compared to a sample of Ibc SNe from \cite{Margutti14b} and \cite{Soderberg10}. SN\,2014C shows an uncommon, steady increase in X-ray and radio luminosities until late times,  a signature of the continued shock interaction with very dense material in the environment. \emph{Upper panels}:  The optical bolometric luminosity of SN\,2014C is well explained at early times by a model where the source of energy is purely provided by the radioactive decay of $^{56}$Ni (grey thick line, top left panel).  However, at later times (top right panel) SN\,2014C shows a significantly flatter temporal decay, due to the contribution of  more efficient conversion of shock kinetic energy into radiation. This evolution is accompanied by a marked increase of H$\alpha$ emission (\emph{Lower Panels}), as a consequence of the SN shock interaction with H-rich material. See M15 and K15 for details about the spectroscopical metamorphosis and the radio evolution, respectively.}
\label{Fig:XRO}
\end{figure*}

\subsection{A continuum of stellar explosions between type Ib/c and type-IIn SNe}
\label{SubSec:Continuum}

\begin{figure*}
\vskip -0.0 true cm
\centering
\includegraphics[scale=0.57]{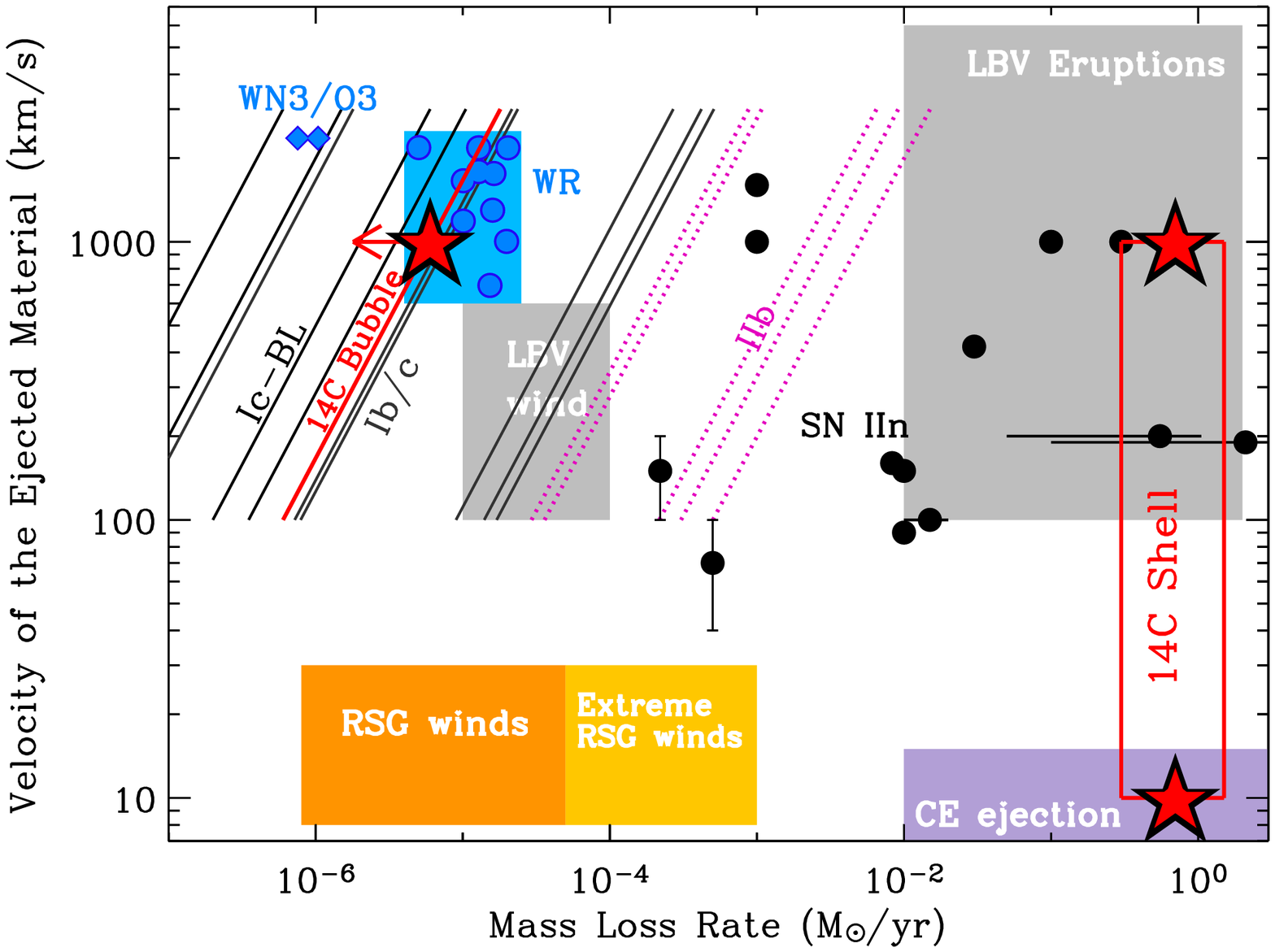}
\caption{The environment sampled by the SN\,2014C shock evolves from the typical low-density environment around Ib/c SNe and WR stars, to the dense and rich environment typical of SNe that develop signatures of strong interaction with the medium (i.e. type-IIn SNe, here represented with black dots, data from \citealt{Kiewe12}). H-poor SNe are represented with diagonal lines since the observations constrain the density $\rho$ which is $ \propto \dot M/v_w$. Black, blue and dotted purple lines are used for the sample of type Ic-BL, Ib/c/ and IIb SNe from \citealt{Drout15}. The properties of galactic WR stars are from \cite{Crowther07}, while WN3/O3 stars are from \cite{Massey15}. Locations of red supergiants environments (RSG) are from \cite{deJager88}, \cite{Marshall04} and \cite{vanLoon05}, while the typical locations of Luminous Blue Variable (LBV) winds and eruptions are inferred from \cite{Smith14} and \cite{Smith06}. For the common envelope (CE) ejection due to binary interaction we use here a typical time scale of 1 yr.}
\label{Fig:continuum}
\end{figure*}

SN\,2014C experienced a remarkable metamorphosis, evolving from an \emph{ordinary} type-Ib SN to a strongly interacting type-IIn SN over the time scale of $\sim1$ yr. Figure \ref{Fig:XRO} captures the key observational  properties of its evolution over the electromagnetic spectrum:  radio, X-ray and H$\alpha$ emission progressively got stronger with time, while the optical light-curve flattened, due to the contribution from the SN shock interaction with dense material in the environment. The strong interaction of the SN shock with a dense, H-rich material is the defining property of SNe of type IIn.  In this sense, SN\,2014C bridges the gap between type Ib/c and type IIn SNe (Fig. \ref{Fig:continuum}). We showed in Sec. \ref{Sec:Env} that this phenomenology requires the deposition of a substantial amount of H-rich material ($\sim1\,\rm{M_{\odot}}$) in the SN surroundings at $R\sim5.5\times 10^{16}\,\rm{cm}$. 

The low density environment at $R\lesssim 2\times 10^{16}\,\rm{cm}$ (Sec. \ref{SubSec:IC}) implies that the progenitor star of SN\,2014C did not suffer massive eruptions within $\sim 7\,(v_w/1000\,\rm{km\,s^{-1}})\,\rm{yrs}$ before death. The location of the dense H-rich shell at $R\sim5.5\times 10^{16}\,\rm{cm}$ argues for an ejection of H-rich material at an earlier epoch corresponding to $\sim20\,(v_H/1000\,\rm{km\,s^{-1}})\,\rm{yrs}$ preceding the explosion, where $v_H$ is the ejection velocity of the shell. For SN\,2014C, the H-rich shell was far enough from the explosion site not to be efficiently ionized by the SN shock and radiation during the first $\sim$100 days -which allowed the early SN classification as an H-poor, type Ib event-, but close enough for the shock-CSM interaction to develop on time scales that are relevant to our coordinated monitoring -which allowed us to witness the later transition to type IIn SN- (Fig. \ref{Fig:continuum}). 

A key difference between ordinary IIn SNe and 14C-like events lies in the location of the H-rich material, which maps into a different epoch of  H-envelope ejection by the stellar progenitors. Ordinary type IIn SNe are characterized by strong interaction with dense CSM since the very first moments after explosion, 
which requires a very recent ejection of the H-rich material (typically within a few years before the stellar
demise) and results in the H-rich material being at $R<10^{16}\,\rm{cm}$ from the explosion center.

Recent observations indeed provided direct evidence for eruptive behavior of progenitors of type IIn-like SNe in the years to days before a major explosion (\citealt{Smith14}). Our observations of SN\,2014C suggest that a fraction of SNe spectroscopically classified of type IIn in fact contained bare type Ib/c-like cores that recently ejected their H envelopes. This suggestion would naturally account for the diverse environments of IIn SNe (which argue for a broad range of progenitor star masses; \citealt{Kelly12}, \citealt{Anderson15}) and it is in line with what was inferred for the type-IIn SN\,1996cr \citep{Dwarkadas10}. In general, our picture would explain the observational properties of type-IIn SNe like 1986J  that did not show evidence for high-velocity H features during its spectroscopic evolution, but showed evidence for broad O   features at late times (\citealt{Rupen87}, \citealt{Bietenholz02}, \citealt{Milisavljevic08}).

Previous studies pointed to a continuum of properties among SNe interacting with He-rich and H-rich material (i.e. type-Ibn SNe and SNe of type IIn,  \citealt{Smith12}, \citealt{Pastorello15a} and references therein). With our study we significantly extend the continuum to include even H-stripped core-collapse explosions. The existence of 14C-like events argues in favor of a continuum of stellar explosions that  bridges the gap between Ib/c and IIn SNe (Fig. \ref{Fig:continuum}) and that directly maps into a continuum of time scales of ejection of the H-rich envelope of progenitors of H-stripped SNe, extending from less than one year to thousands of years before explosion. 

\subsection{The origin of the H-rich shell in SN\,2014C}
\label{SubSec:CMSorigin}
SN\,2014C is an ordinary type Ib/c SN embedded in a non-standard H-rich environment shaped by the mass-loss history of its progenitor system. Here we discuss the origin of the H-rich, dense CSM material around SN\,2014C in the context of the stellar evolution and mass-loss mechanisms of evolved massive stars. 

Single massive stars that are progenitors of ordinary H-stripped SNe are expected to evolve from H-rich, extended Red SuperGiants (RSGs) into core-He-burning compact Wolf-Rayet (WR) stars  $\sim0.5-1$ Myr before exploding as H-poor SNe (e.g. \citealt{Heger03}). The presence of H-rich material within $\sim6\times 10^{16}\,\rm{cm}$ of the explosion site of SN\,2014C demands an exceptionally short WR phase of its progenitor star, lasting only $\sim20$ yrs (for an ejection velocity of the H-rich material $v_H=1000\,\rm{km\,s^{-1}}$),  and a very large mass-loss rate during the previous RSG phase $\dot M_{RSG}\geq 2\times 10^{-3}\,\rm{M_{\sun}yr^{-1}}$ (the inequality accounts for the fact that the RSG wind is in fact freely expanding).  This value is $\geq10$ times larger than the typically observed mass-loss rates during the RSG phase (observations indicate $\dot M_{RSG}<10^{-4}\,\rm{M_{\sun}yr^{-1}}$,  \citealt{vanLoon05}, \citealt{Mauron11}) and exceeds the limit for mass loss due to the commonly assumed line-driven winds even in the case of a small filling factor $f=0.1$ of the H-rich shell (e.g. \citealt{Smith14}, their Fig. 3).\footnote{Another possibility is that SN\,2014C exploded in an environment enriched by its stellar companion. This scenario would require having an ordinary SN Ib/c exploding within $\sim20$ yrs of a giant LBV-like eruption from its stellar companion by chance. Non-terminal, giant eruptions are typically associated to very massive stars (i.e. Luminous Blue Variable stars, LBV, \citealt{Humphreys94}), which would require the progenitor of SN\,2014C to be even more massive in order to exhaust its H fuel and explode before its companion. The standard explosion properties of SN\,2014C and its ordinary luminosity do not support this scenario (Sec. \ref{SubSec:ExplPar} and Fig. \ref{fig:Lbol}). Furthermore, this picture is unlikely to be able to account for the entire fraction of $\sim10$\% of SNe Ib/c that present a 14C-like behavior (Sec. \ref{SubSec:rates14C}). }  

We thus conclude that SN\,2014C violates several expectations from standard evolutionary models that employ time averaged mass-loss prescriptions and do not include highly time-dependent mass-loss episodes associated with instabilities during the last stages of stellar evolution and the effects of binarity. With SN\,2014C we further have clear evidence that mass-loss mechanisms other than the commonly-assumed metallicity-dependent line-driven winds
are active and play an important role  in the process that leads to the H-stripped progenitors of SNe Ib/c. This picture, if common among Ib/c SN progenitors,  naturally explains why the mass-loss rates inferred for the progenitors of SNe Ibc span a much larger range than the observed wind mass-loss rates of WRs in our Galaxy  (e.g. \citealt{Soderberg07}, \citealt{Wellons12}). Finally, the indication of non-metallicity driven mass-loss mechanisms being at play in the evolutionary path that leads to Ib/c SNe is  also consistent with inferences from the demographics of SN types combined with initial-mass function considerations \citep{Smith11}. 

To explain the observations of SN\,2014C we are forced to leave behind the traditional view of single massive star evolution and explore the possibility of alternative  physical mechanisms responsible for the ejection of the H envelope of its progenitor, with the following key requirements:
\begin{itemize}
\item[(1)] Synchronized with the explosion.
\item[(2)] Efficient, and able to (almost) entirely strip the star of its last H layer.
\item[(3)] Common to 10\% of progenitors of H-stripped SNe during their last $\sim500$ years of evolution, but potentially to every SN Ib/c progenitor \emph{if} the process is intrinsically active over a longer time scale of $5000$ yrs or more.
\end{itemize}

We consider below the effects of binary evolution and late-stage nuclear burning instabilities, which are typically not incorporated in the current prescriptions of mass loss in evolved massive stars. 
\subsubsection{Binary Interaction}
\label{SubSubSec:binary}

The ejection of the H-rich envelope can be the result of binary interaction (e.g. \citealt{Podsiadlowski92}). Since the majority of young massive stars are found in binary systems that are close enough to interact (e.g. \citealt{Sana12}), these ejections are expected to be common, but not necessarily synchronized  with the stellar death. 

The binary interaction can strip the star of almost all of its H, leaving behind just a thin H layer on its surface (which would be later lost through metallicity-dependent, line-driven winds). The envelope material 
is expected to get compressed into a thin and dense shell, consistent with what we observe for SN\,2014C. Furthermore, a binary progenitor for SN\,2014C is also suggested by the pre-explosion observations of the cluster of stars that hosts SN\,2014C, which favor lower mass star progenitors with $M<20\,\rm{M_{\sun}}$ (M15). Stars with relatively lower masses are disfavored as progenitors of Ib/c SNe in a single-star progenitor scenario because of their weaker winds, which are not able to entirely strip the star of its H-envelope. Stars with relatively lower masses  are instead the most common progenitors of Ib/c SNe in a binary progenitor scenario, where the interaction with the companion would bear the burden of the envelope removal.
Finally, since the observed fraction of $\sim37$\%  of H-poor SNe among core-collapse explosions is too large to be reconciled with the evolution of single massive stars (e.g. \citealt{Kobulnicky07}, \citealt{Eldridge08}, \citealt{Li11}, \citealt{Smith11}), it is expected that at least a fraction of SNe Ibc originate in binaries. 

However, in general we do not expect the stellar death and the binary interaction to be related events that are synchronized in time.  The time lag between the interaction and stellar death can be small if the progenitor fills its Roche lobe in a late evolutionary stage. This occurs for systems with orbital periods longer than $\sim$$1000$ days that undergo case-C mass transfer (e. g. Fig.~1 in  \citealt{Schneider15}). In such systems the progenitor fills its Roche lobe as a red super giant, after finishing central helium burning.  At this stage the star typically has less than a few thousand years (depending on its mass) to live until its final explosion. The convective envelope of the giant responds to mass loss by expanding, leading to  dynamically unstable mass transfer in which the companion is engulfed in the envelope of the red supergiant. The envelope is ejected as the companion spirals in, which takes a few orbits at most (\citealt{Ivanova13}).  This means that  several solar masses of hydrogen rich material are ejected within a short time ($\lesssim 1$ year), leading to a dense shell, torus or disk around the binary system.  

The outflow velocities of the H-rich material are expected to be comparable  to the escape speed of the surface of the original red supergiant (up to a few $10$ km\,s$^{-1}$), or possibly even less if the material remains bound to the system and resides in a circum-binary disk. It is unclear how long the ejected material can survive in the vicinity of the system. The ejected material will be shaped and eventually eroded by the fast line-driven wind of the stripped progenitor and its ionizing photons \citep{Metzger10}. 

Case-C mass transfer is relatively rare for binary systems where the primary is massive enough to produce a SN. \cite{Podsiadlowski92} estimate this fraction to be $5.6\%$ among binary systems with primaries with mass in the range $8-20\,\rm{M_{\sun}}$.
For a more accurate estimate of this fraction and to obtain the distribution of time lags
we use the binary star population synthesis code binary\_c (\citealt{Izzard09,Izzard06,Izzard04,deMink13}) which relies on the algorithms by \cite{Hurley00, Hurley02}. The crucial ingredients in this estimate are the stellar lifetimes and radial expansion of the stars, which follow from the detailed stellar evolutionary models by \cite{Pols98}.  These algorithms provide a computationally efficient approximation that allows to simulate a full population of binary systems. The approximations are sufficient for the scope of this work, but future investigation with detailed codes would be desirable.   

We simulate binary populations by choosing primary masses from a \cite{Kroupa01} mass function and companion masses uniformly distributed with a mass ratio between 0.1 and 1 \citep[e.g.][]{Duchene13, Sana12}. We assume a uniform distribution of initial periods $p$ in log space (\"Opik's law). However, for systems with initial masses above $15\,\rm{M_{\sun}}$ we adopt the distribution obtained by \citet{Sana12} that more strongly favors short period systems, with $ 0.15 \le \log_{10} p \le 3.5$.  We assume a binary fraction of 0.7.  
We account for the relevant physical processes that govern binary systems including stellar wind mass loss, Roche-lobe overflow and common envelope (CE) phases using the standard assumptions for the physics parameters as summarized in \citet{deMink13} and references therein. In addition, apart from our standard model we also run simulations with different assumptions for the initial conditions (i.e. slope of the initial mass function, binary fraction, distribution of mass ratios and orbital periods) and the treatment of the physical processes (i. e. efficiency of common envelope ejection, efficiency of mass transfer, critical mass ratio for the onset of contact, amount of mixing and mass loss assumed during a stellar merger). We find that the variations in the initial mass function dominate the uncertainty. We refer to \cite{deMink13}, \cite{deMink14} and \cite{Izzard04} for a detailed discussion. The full set of simulations will be published in Zapartas et al. (in prep).

We find that $\sim$6.5\% (3.8 -- 10\%) of type Ibc SN progenitors result from systems that experienced case-C CE evolution. The 3.8 -- 10\% range reflects a variation of the slope of the initial mass function $\alpha_{\rm{IMF}}=-2.3\pm0.7$. The other parameters of our model  induce smaller variations to the final fraction of progenitors that went through case-C CE evolution. The fraction of progenitors with recent CE evolution is thus somewhat smaller than -but reasonably close to- our estimates of interacting systems of Sec. \ref{Sec:Rates}. The typical progenitors are stars of $8-20\,\rm{M_{\sun}}$ in wide binary systems of $\gtrsim 1000$ days, which experience unstable mass transfer and CE after their core He exhaustion. At this stage, the life time before core collapse is similar to the time scale for carbon burning, which is typically a few $10^3$ yrs, but potentially extends to a few  $10^4$ yrs for the lowest-mass progenitors (\citealt{Jones13}). This fact implies that binary interaction can be the main culprit \emph{if} the material ejected during the CE phase is able to survive close to the system for a few $\sim10^3-10^4$ yrs.


Alternatively we need to invoke a new physical mechanism or more exotic channels to enhance the fraction of Ibc SNe with CE interaction shortly before the explosion.  A possibility is to consider exotic binary evolutionary channels where the final explosion has a causal relation with the binary interaction.  One example is the reverse CE inspiral of a neutron star or black hole in the envelope of the companion as discussed by \cite{Chevalier12}. Based on the estimates by \cite{Chevalier12} this channel would be able to account for 1-3\% Ibc  SNe.
Alternatively, the fraction of interacting systems can be increased if the stellar radius of the progenitor inflates before the explosion.  Instabilities during the last phases of nuclear burning evolution might inject the necessary energy that leads to the envelope inflation (Smith \& Arnett 2014), and might explain the larger fraction of interacting Ibc supernovae in our sample. 

\subsubsection{Instabilities during the final Nuclear Burning Stages}
\label{SubSubSec:burning}
Instabilities associated with the final nuclear burning sequences (especially O and Ne) have been recently invoked to explain the observation of eruptions from massive stars in the months to years before core collapse (\citealt{Arnett11}, \citealt{Quataert12}, \citealt{Shiode13}, \citealt{Shiode13b}, \citealt{Smith14b}).   
This mechanism is naturally synchronized with the stellar demise and it is expected to be common to most stars. However, nuclear burning instabilities do not necessarily lead to the almost complete stripping of the H layer that we observe in SN\,2014C.  Even more importantly, our observations of SN\,2014C require the H envelope ejection to have happened $\geq 20$ years before the explosion, definitely \emph{before} the start of the O-burning phase.
Heavy mass loss in the H-stripped SN\,2014C was thus not limited to the few years preceding the collapse, and, if connected to nuclear burning instabilities, directly points to the development of instabilities even at earlier times in the nuclear burning sequence.

Current theoretical investigations have explored  with realistic simulations only the very late nuclear burning stages (O, Ne and later stages, with duration of the order of $\sim$ yrs) mainly because of limitations in computational power (\citealt{Meakin06}, \citealt{Arnett11b}, \citealt{Smith14b}, \citealt{Smith14}). SN\,2014C and our analysis of type Ib/c SNe of Sec. \ref{SubSec:Stat} is however suggestive of a significantly longer active time  of the physical process behind the ejection of massive H-rich material. For $v_{H}\leq1000\,\rm{km\,s^{-1}}$, $\Delta t_{\rm{active}}\geq5000$ yrs, thus extending well back into the C-burning stage of massive stars (e.g. \citealt{Yoon10,Shiode14}). 


 It is thus urgent to theoretically explore the possibility of instabilities during  the earlier stages of nuclear burning, potentially extending to C-burning.\footnote{While the exact time scales of each burning stage is sensitive to the currently imposed mass-loss rates in stellar evolution models which do not include the effects of time-dependent mass loss discussed in this section, we note that the detection of infrared echoes from distant shells in the environments of IIn (e.g. \citealt{Fox11}) and non-IIn SNe (an illustrative example is the ring of material at $\sim6\times10^{17}$ cm from the explosion site of SN\,1987A, \citealt{Sonneborn98}) independently supports the idea that enhanced mass loss is not confined to the few years before the stellar collapse. }  The -so far- neglected time dependence of nuclear burning and mass loss in massive stars might have a fundamental influence on the pre-SN structure of the progenitor star, a key input parameter to all numerical simulations of SN explosions \citep{Janka12}. 

\section{Summary and Conclusions}
\label{Sec:conc}
SN\,2014C represents the first case of a SN originating from an H-stripped progenitor for which we have been able to closely monitor a complete metamorphosis from an ordinary Ib-SN into a strongly interacting type-IIn SN over a time scale of $\sim1$ yr \citep{Milisavljevic15}. Observational signatures of this evolution appear across the electromagnetic spectrum, from the hard X-rays to the radio band.  The major finding from our study  of SN\,2014C is the  presence of substantial ($M\sim1\,M_{\odot}$) H-rich material located at $R\sim6\times10^{16}\,\rm{cm}$
from the explosion site of an H-poor core-collapse SN. This phenomenon challenges the current theories of massive stellar evolution and argues for a revision of our understanding of mass loss in evolved massive stars. Specifically:
\begin{itemize}
\item With $E_{\rm{k}}\sim1.8\times10^{51}\,\rm{erg}$, $M_{\rm{ej}}\sim1.7\,\rm{M_{\sun}}$ and $M_{\rm{Ni}}\sim0.15\,\rm{M_{\sun}}$, the explosion parameters of SN\,2014C are un-exceptional among the population of Ib/c SNe. 
\item SN\,2014C adds to the complex picture of mass loss in massive stars that recent observations are painting (\citealt{Smith14}) and demonstrates that the ejection of massive H-rich material is not a prerogative of very massive H-rich stars ($M\sim60\,\rm{M_{\sun}}$, like the progenitor of SN\,2009ip, \citealt{Smith10},  \citealt{Foley11b}). Instead it shows that even progenitors of normal H-poor SNe can experience severe pre-SN mass loss as late as $10\lesssim t\lesssim$ 1000 years before explosion. Heavy mass loss in SNe Ib/c is thus not limited to the few years preceding core collapse.
\item In this sense SN\,2014C  bridges the gap between ordinary SNe Ib/c and type-IIn SNe, which show signs of shock interaction with a dense medium from the very beginning. The existence of 14C-like events establishes a continuum of time-scales of ejection of substantial H-rich material by massive stars, extending from $<1$ yr  before collapse for type-IIn SNe, to decades and centuries before explosion for Ib/c SNe. This fact leads to the idea that a fraction of spectroscopically classified type-IIn SNe in fact harbor bare Ib/c-like cores that underwent a very recent ejection of their H-rich envelopes.
\item SN\,2014C violates the expectations from the standard metallicity-dependent line-driven mass loss channel and demonstrates the existence of a time-dependent  mass loss mechanism that is active during the last centuries of evolution of some massive stars and that leads to progenitors of ordinary H-poor core-collapse SNe. Possibilities include the effects of the interaction with a binary companion or instabilities during the last nuclear burning stages. In both cases we do not expect a strong metallicity dependence.
\item We analyzed 183 Ib/c SNe with radio observations and we found that 10\% of SNe in our sample displays evidence for late-time interaction reminiscent of  SN\,2014C. This fraction is somewhat larger than -but in reasonable agreement with- the expected outcome from recent envelope ejection due to binary evolution \emph{assuming} that the envelope material can survive close to the progenitor site for $10^3-10^4$ yrs. Alternatively, events related to the last phases of nuclear burning might also play a critical role providing the energy and the trigger mechanism that cause the ejection of envelope material from an evolved massive star.
\item In particular, our analysis suggests that unsteady nuclear burning (i) may be spread across a wide range of initial progenitor mass to include not only the most massive stars, but also ordinary progenitors of Ib/c SNe; (ii) instabilities are not confined to the O, Ne and Si-burning phases, but  instead likely extend all the way to C-burning; (iii) unsteady nuclear burning might enhance the fraction of binary interactions before collapse. 
\end{itemize}

The findings and conclusions above highlight the important role of time-dependent, eruptive mass loss in the evolutionary path that leads to the progenitors of ordinary H-poor core-collapse SNe. \emph{The incorrect use of time averaged mass-loss prescriptions in current models of stellar evolution might have a major effect on our understanding of the stellar structure of a massive stellar progenitor approaching core collapse and might lead to inaccuracies in pre-SN stellar structure that are of fundamental importance at the time of the explosion.} To make progress  it is urgent to theoretically explore the presence of instabilities during the earlier stages of nuclear evolution in massive stars, and in general, to study the effects of significant eruptive mass loss on the pre-supernova stellar structure. Observationally, it is mandatory to consistently sample the pre-SN life of stellar progenitors in the \emph{centuries} before explosion, a territory that can only be probed with late-time radio and X-ray observations of nearby stellar explosions.





\acknowledgments 
R. M. acknowledges generous support from the James Arthur Fellowship at NYU.  S.~d.~M acknowledges support by a Marie Sklodowska-Curie Reintegration Fellowship (H2020 MSCA-IF-2014, project id 661502). M.~Z. acknowledges support by the Netherlands Research School for Astronomy (NOVA). The National Radio Astronomy Observatory is a facility of the National Science Foundation operated under cooperative agreement by Associated Universities, Inc. The scientific results reported in this article are based on observations made by the Chandra X-ray Observatory under programs GO 15500831 and DDT 15508491. This work was partially supported under NASA No. NNX15AV38G, and made use of data from the Nuclear Spectroscopic Array (NuSTAR) mission, a project led by Caltech, managed by the Jet Propulsion Laboratory, and funded by the National Aeronautics and Space Administration. This work was supported in part by National Science Foundation Grant No. PHYS-1066293 and the hospitality of the Aspen Center for Physics. We thank the Chandra, NuSTAR and Swift teams for support with the execution of the observations. 
\appendix
\section{Tables}

\begin{deluxetable*}{cccccccccccc}[b!]
\tabletypesize{\scriptsize}
\tablecolumns{9} 
\tablewidth{0pc}
\tablecaption{\emph{Swift}-UVOT photometry.}
\tablehead{ \colhead{Date} & \colhead{$v$} & \colhead{Date} & \colhead{$b$} & \colhead{Date} &  \colhead{$u$} & \colhead{Date} &  \colhead{$w1$} & \colhead{Date} &  \colhead{$w2$ }& \colhead{Date} &  \colhead{$m2$}
\\ \colhead{(d)}  & \colhead{(mag)} & \colhead{(d)} & \colhead{(mag)} & \colhead{(d)} &  \colhead{(mag)} & \colhead{(d)} &  \colhead{(mag)} & \colhead{(d)} &  \colhead{(mag) }& \colhead{(d)} &  \colhead{(mag)}}
\startdata
663.25\footnote{Dates are in MJD-56000 (days).} & 15.29(0.06)\footnote{Not host subtracted. Not extinction corrected. Uncertainties are 1$\sigma$.} & 663.25 & 16.35(0.06) & 663.25 & 16.46(0.08) & 663.28 & 17.84(0.13) & 663.25 & >18.68 & 663.26 & >18.93 \\ 
664.18 & 15.18(0.07) & 664.21 & 16.17(0.06) & 664.21 & 16.54(0.08) & 664.21 & 17.69(0.10) & 664.21 & >18.93 & 664.22 & >19.12 \\ 
664.25 & 15.15(0.06) & 666.45 & 15.85(0.05) & 666.45 & 16.34(0.07) & 666.59 & 17.73(0.07) & 960.21 & >19.10 & 960.21 & >19.03 \\ 
666.45 & 14.79(0.04) & 666.52 & 15.81(0.05) & 668.45 & 16.46(0.07) & 668.59 & 17.76(0.07) & 1013.69 & >19.15 &  &  \\ 
666.52 & 14.80(0.04) & 668.52 & 15.81(0.05) & 668.52 & 16.41(0.07) & 670.39 & 17.80(0.07) &   &  &  \\ 
668.52 & 14.68(0.06) & 670.45 & 15.81(0.05) & 670.45 & 16.66(0.08) & 672.46 & 17.88(0.07) &  &  &  & \\ 
670.45 & 14.69(0.04) & 670.52 & 15.83(0.05) & 672.52 & 16.63(0.08) & 674.73 & 18.00(0.08) &  &  &  &  \\ 
670.52 & 14.67(0.04) & 672.39 & 15.91(0.05) & 674.79 & 16.75(0.08) & 676.32 & 17.94(0.07) &  &  &  & \\ 
672.39 & 14.71(0.04) & 672.52 & 15.87(0.05) & 676.06 & 16.83(0.08) & 960.21 & >18.27 &  &  &  &  \\ 
672.52 & 14.69(0.04) & 674.12 & 15.96(0.05) & 676.59 & 16.76(0.08) &  &  &  &  &  &  \\ 
674.12 & 14.77(0.04) & 674.80 & 16.02(0.05) &1044.80& >17.25&  &  &  &  &  &  \\ 
674.80 & 14.78(0.04) & 676.06 & 16.11(0.05) &  &  &  &  &  &  &  &  \\ 
676.06 & 14.83(0.04) & 676.59 & 16.04(0.05) &  & &  &  &  &  &  &  \\ 
676.59 & 14.80(0.04) &  &  &  & &   &   &  &  \\ 
\enddata
\label{Tab:UVOTphot}
\end{deluxetable*}

\begin{deluxetable}{cccc}[b!]
\tabletypesize{\scriptsize}
\tablecolumns{4} 
\tablewidth{10pc}
\tablecaption{MMTCam photometry.}
\tablehead{  \colhead{Date} & \colhead{$r$} & \colhead{$i$}}
\startdata
795 \footnote{Dates are in MJD-56000 (days).} & 17.34(0.01)\footnote{Not host subtracted. Not extinction corrected. Uncertainties are 1$\sigma$.} & 17.05(0.02)\\
977 & 18.86(0.05)& 18.89(0.08) \\
1132 & 19.64(0.09)& 19.57(0.13)\\
1162 & 19.60(0.09) & 19.65(0.14)\\
1165 & 19.50(0.11) & 19.53(0.14)\\
\enddata
\label{Tab:MMT}
\end{deluxetable}

\begin{deluxetable}{cccccccccc}[b!]
\tabletypesize{\scriptsize}
\tablecolumns{10} 
\tablewidth{30pc}
\tablecaption{Sample of 183 Ib/c SNe with radio observations}
\tablehead{ }
\startdata
1954A&1983N&1984L&1985F&1990B&1990U&1991A&1991N&1991ar\\
1994I&1994ai&1996D&1996aq&1997B&1997C&1997X&1997dc&1998bw\\
1998T&1999bc&1999di&1999dn&1999ec&1999eh&1999ex&2000C&2005C\\
2000F&2000H&2000S&2000ds&2000dv&2000cr&2000ew&2000fn&2001B\\
2001M&2001ai&2001bb&2001ch&2001ci&2001ef&2001ej&2001em&2001is\\
2002J&2002ap&2002bl&2002bm&2002cj&2002cp&2002dg&2002dn&2002ge\\
2002gy&2002hf&2002hn&2002ho&2002hy&2002hz&2002ji&2002jj&2002jp\\
2002jz&2003I&2003L&2003ih&2003kb&2003jg&2003jd&2003is&2003ig\\
2003hp&2003id&2003gk&2003gf&2003ev&2003el&2003ds&2003dr&2003dg\\
2003cr&2003bu&2003bp&2003bm&2003aa&2003I&2003H&2003A&2004ao\\
2004ax&2004aw&2004bf&2004bi&2004bm&2004bs&2004bw&2004bu&2004cc\\
2004dc&2004dk&2004dn&2004dx&2004eh&2004eu&2004fe&2004ff&2004ge\\
2004gk&2004gq&2004gt&2004gv&2005E&2005C&2005V&2005N&2005O\\
2005aj&2005ar&2005az&2005bf&2005bh&2005bj&2005bk&2005bq&2005ce\\
2005ct&2005da&2005cz&2005dg&2005ek&2005eo&2005hg&2005ke&2005kl\\
2005kf&2005kz&2005la&2005lr&2005mf&2005nb&2006F&2006ab&2006bf\\
2006bk&2006cb&2006ck&2006dj&2006dl&2006dg&2006dn&2006ea&2006eg\\
2006ei&2006ec&2006ep&2006el&2006fo&2006gi&2006jc&2006lt&2007C\\
2007D&2007I&2007Y&2007bg&2007cl&2007gr&2007iq&2007ke&2007ru\\
2007rz&2007uy&2008D&2008du&2008dv&2009bb&2010ay&PTF11qcj&2012ap\\
2012au&2013ge&2014ad\\
\enddata
\label{Tab:radiosample}
\end{deluxetable}
\bibliographystyle{apj}

\end{document}